\documentclass[12pt,preprint,longabstract]{aastex}

\shorttitle{L. \v C. Popovi\' c et al.}
\shortauthors{Microlensing of the Fe K$\alpha$ line and the X-ray continuum}

\begin{document}

\title{A study of the correlation between the amplification of
the  Fe K$\alpha$ line
and the X-ray continuum of quasars due to microlensing}

\author{L. \v C. Popovi\' c\altaffilmark{1,2},
P. Jovanovi\' c\altaffilmark{1,2}, E. Mediavilla\altaffilmark{3}, A.F.
Zakharov\altaffilmark{4,5,6}, C. Abajas\altaffilmark{3}, J. A.
Mu\~noz\altaffilmark{3,7}, G. Chartas\altaffilmark{8} }
\altaffiltext{}{
$^1$Astronomical Observatory, Volgina 7, 11160 Belgrade
74, Serbia \\
$^2$Isaac Newton Institute of Chile, Yugoslavia Branch and Universidad Diego Portales, Chile\\
$^3$Instituto de Astrofisica de Canarias, 382005 La Laguna, Tenerife,
Spain\\
$^4$Institute of Theoretical and Experimental Physics,
25, B.Cheremushkinskaya st., Moscow, 117259, Russia \\
$^5$Astro Space Centre of Lebedev Physics Institute, Moscow, Russia\\
$^6$Isaac Newton Institute of Chile, Moscow Branch\\
$^7$Departamento de Astronom\'{\i}a y Astrof\'{\i}sica, Universidad
de Valencia, E-46100 Burjassot, Valencia, Spain\\
$^8$Astronomy and Astrophysics Department, Pennsylvania State University,
University Park, PA 16802, USA}

\begin{abstract}
The observed enhancement of the Fe K$\alpha$ line in three
gravitationally lensed QSOs (MG J0414+0534, QSO 2237+0305,
 H1413+117) is interpreted in terms of microlensing, even when equivalent
X-ray continuum amplification is not observed. In order to
interpret these observations, first we studied
the effects of microlensing on quasars spectra, produced by
straight fold caustic crossing over standard relativistic accretion disk.
The disk emission was analyzed using the ray tracing method,
considering Schwarzschild and Kerr metrics.
 When the emission is separated in two regions (an inner disk corresponding to
the Fe K$\alpha$ line and an outer annulus corresponding to the
 continuum, or vice-versa) we find microlensing events which enhance the
Fe K$\alpha$ line without noticeable
  amplification of the X-ray continuum, but only during a limited time
 interval. Continuum amplification is expected if a complete microlensing
event is monitored. Second, we studied a more realistic case of
amplification by caustic magnification pattern.
In this case we could satisfactorily explain
the observations  if the Fe K$\alpha$ line is emitted from the
innermost part of the accretion disk, while the continuum is emitted from
a larger region.  Also, we studied the chromatic effects of
microlensing, finding that the
 radial distribution of temperature in the accretion disk, combined with
 microlensing itself, can induce wavelength dependent variability of $\sim$ 30\%
for microlenses with very small masses.
  All these results show that X-ray
monitoring of gravitational lenses
is a well suited method for studying of the innermost
 structure of AGN accretion disks.
\end{abstract}


\section{Introduction}

Recent observational and theoretical studies suggest that gravitational
microlensing can induce variability in the X-ray emission of
lensed QSOs. Microlensing of the Fe K$\alpha$ line
has been reported at least in three macrolensed QSOs:  MG
J0414+0534 \citep{Chart02a},  QSO 2237+0305 \citep{Dai03}, and H1413+117
\citep{Osh01, Pop03, Chart04}.

The influence of microlensing in the X-ray emission has been also
theoretically investigated. \cite{Min01} simulated the variation of the
X-ray continuum due to microlensing showing that the flux magnifications
for the X-ray and optical continuum emission regions are not
significantly different during the microlensing event, while
\cite{Yon98,Yon99,Tak01} found that simulated spectral variations caused
by
microlensing show different behaviour, depending on photon energy. Also,
microlensed light curves for thin accretion disks around Schwarzschild and
Kerr black holes were considered in \cite{Jar92} and microlensing light
curves for the Fe K$\alpha$ were simulated by \cite{Jar02}. On the other
hand, the influence of microlensing in the Fe K$\alpha$ spectral line
shape was discussed in \cite{Popov01,Chart02a} and
\cite{Pop03,Pop03a}\footnote{Simulations of X-ray line profiles are
presented in a number of papers, see, for example,
 \cite{Fabian01,Zak_rep02,Zak_rep02a,Zak_rep02_xeus} and references
therein, in particular \cite{ZKLR02} showed
that information about the magnetic filed may be extracted from X-ray
line-shape analysis; \cite{Zak_rep03a_AA,Zak_rep03b_AA} discussed
signatures of X-ray line-shapes for highly inclined accretion
disks.}. \cite{Pop03,Pop03a} showed that objects in a foreground galaxy with even relatively small
masses can produce observable changes in the Fe K$\alpha$ line flux,
much stronger than those expected for the UV and optical lines
 (\cite{Pop01b,Aba02,li04}).
In the optical spectra, microlensing induced magnification of
broad UV lines (e.g., CIV and SIV/OIV)
was reported by Richards et al.  (2004).  Consequently, one
can expect that
microlensing of the Fe K$\alpha$ line region can be more frequent.
Observations of the X-ray continuum and
the Fe K$\alpha$ line in multi-imaged AGNs open new possibilities
for the study of the unresolved X-ray emitting structure in QSOs,
particularly
for high redshifted  QSOs  \citep{Zak03,Dai04}.

However, an explanation for different behavior of line and continuum
variability in the observed events
should be given in context of the microlensing hypothesis.
\cite{Chart02a}  detected an
increase of the Fe K$\alpha$ equivalent width in the image B of the lensed
QSO J0414+0534 that was not followed by the continuum. \cite{Chart02a}
explained the non-enhancement of the continuum emission in the spectrum of image B
by proposing that the thermal emission region of the disk and the
Compton up-scattered emission region of the hard X-ray source lie within
 smaller radii than the iron-line reprocessing region. Analyzing the
X-ray variability of QSO 2237+0305A,
\cite{Dai03} also measured amplification of
the Fe K$\alpha$ line in component A of QSO 2237+0305 but not in
 the continuum. However, in this
case the interpretation was different. \cite{Dai03} suggested that the
larger size of the continuum emission region ($\sim 10^{14}$ cm $\sim$ 100 R$_g$ for
M$_{BH}=10^7M_\odot$) with respect to the Fe K$\alpha$ emission
 region ($\sim$ 10 R$_g$) could explain this result.
Finally, in H~1413+117 \cite{Chart04} found that the continuum and the Fe K$\alpha$
line were enhanced by a different factor.

 With the aim of discussing these results, we will model here the behavior
of the X-ray continuum and the Fe K$\alpha$ line during a
 microlensing event for different sizes of the continuum and the Fe
 K$\alpha$ line emission regions.

\section{Microlensing of a compact accretion disk}

\subsection{The model}

The assumption of a disk geometry for the distribution of  the Fe
K$\alpha$ emitters is supported by the spectral shape of this line in
AGN (e.g.
\cite{Nan97}, where they have investigated the iron line
properties of 18  Sy 1 galaxies). Regarding the
X-ray continuum emission, it seems
that it mainly arises from an accretion disk. For instance,
\cite{Fab03} have shown that the X-ray spectral variability of
MCG-6-30-15 can be modeled  by a two-component model where the one varying
 component is a power-law and the other constant component is produced
by very strong reflection from a relativistic disk.

To study the effects of microlensing on a compact accretion disk we will
use the ray tracing method considering only those photon
trajectories that reach the sky plane at a given observer's angle
$\theta_{\rm  obs}$ (see e.g. \cite{Pop03} and references therein). The
amplified  brightness with amplification $A(X,Y)$ for the
continuum is given by

\begin{equation}
I_{C} (X,Y;E_{obs}) = { {I_{P}}}  (E_{obs},T(X,Y)) \cdot A(X,Y),
\end{equation}
 and for the Fe K$\alpha$ line by

\begin{equation}
\begin{array}{ll}
I_{L} (X,Y;E_{obs}) = & { {I_{P}}}  (E_{0} \cdot g(X,Y),T(X,Y)) \cdot \\
               &\delta (E_{obs}-E_{0} \cdot g(X,Y)) \cdot A(X,Y), \nonumber
\end{array}
\end{equation}
where  $T(X,Y)$ is the temperature, $X$ and
$Y$ are the impact parameters which describe the apparent position of each
point
of the accretion disk image on the celestial sphere as seen by an observer
at infinity;  $E_{0}$ is the the line transition  energy
($E_{0}^{\rm Fe\ K\alpha}=6.4$ keV) and {$g(X,Y)=E_{\rm obs}/E_{\rm em}$ is the energy shift due to relativistic effects ($E_{\rm obs}$ is the observed energy and $E_{\rm em}$ is the emitted energy from the disk). Here we will not consider the cosmological redshift.}  The emissivity of the disk is one of the
important parameters which has influence on the continuum and line
shapes. The
observed continuum flux is very often fitted with one or two black-body
components in the soft X-ray, in addition to the hard X-ray power
law (see e.g. Page et al. 2004).
The line shape, as well as the continuum distribution, strongly
depend on emissivity law.
In the standard Shakura-Sunyaev disc model \citep{Shakura73},
accretion occurs via an optically thick and geometrically thin disc. The
effective optical depth in the disc is very high and photons are close to
thermal equilibrium with electrons. The surface temperature is a function
of disk parameters and results in the multicolor black body spectrum.
This component is thought to explain the
'blue bump'
in AGN and the soft X-ray emission in galactic black holes.
Although the standard model
does not predict the power-law X-ray emission observed in all
sub-Eddington accreting black holes, the power law for the X-ray
emissivity in AGN is usually accepted (see e.g. Nandra et al.
1999). But  one can not   exclude other emissivity laws,
such as black-body or modified black-body emissivity laws.
 Therefore, we will use here black-body, modified
black-body and power emissivity laws for both; the Fe K$\alpha$ and
continuum emission.

In the case of the black-body radiation law, the disk emissivity  is
given as (e.g. Ja\-ro\-szy\'n\-ski et al. 1992):

$$I_P(X,Y;E)=B[E,T_s(X,Y)],$$
where
\begin{equation}
 B\left( {E,T_s(X,Y)} \right) = {\frac{{2E
^{3}}}{{h^2c^{2}}}}{\frac{{1}}{{e^{{{E
}
\mathord{\left/ {\vphantom {{h\nu}  {kT}}} \right.
\kern-\nulldelimiterspace} {kT_s(X,Y)}}} - 1}}},
\end{equation}
where $c$ is the speed of light, $h$ is the Planck constant, $k$ is the
Boltzmann constant and $T_s(X, Y)$ is the  surface temperature
of X-ray accretion disk.

In principle, one can assume different distribution of the
surface temperature along disk. To obtained the X-ray continuum
distribution using Eq. (3) one can assume that
$T_s=const.$, taking that the black hole is powerful  X-ray
sources  with an effective temperature 10$^7$ to 10$^8$ K. But, regarding
the standard disk model it is
expected that the surface temperature  at least is
radially dependent.
Therefore,
here we will accept the radial distribution of surface temperature
given by \citep{Shakura73}:
 \begin{equation}
T_s(X, Y) \sim r^{-3/2}(X,Y)(1-r^{-1/2}(X,Y))^{4/5} \,{\rm K},
\end{equation}
taking that an effective temperature is in an interval from 10$^7$ to
10$^8$
K.
The distribution of the temperature along the radius of the disk used in
 this paper is given in Fig. 1 (top) and corresponding shape of spectral
energy distribution is shown in Fig. 1 (bottom).
In Eq. (4) $r$ is the dimensionless parameter
defined as:
\begin{eqnarray}
r(X,Y)=\frac{R(X,Y)}{6 R_g}=
\frac{1}{6}\frac{R(X,Y)c^2}{GM}=
\frac{M_\odot}{M}\frac{R(X,Y)}{9~{\rm km}},\nonumber
\end{eqnarray}
\noindent
where $R(X,Y)$ is disk radius, expressed in gravitational radii $R_g$.

However, in the innermost part of the accretion disk the Planck function
cannot be used properly.
Therefore we will  use also the standard (classical) Shakura -- Sunyaev
approach, where
the emissivity law is
described by a "modified" black-body radiation law (Eqs. (3.4),
(3.8) in \cite{Shakura73}; see also the discussion in
\cite{Novikov73,Shapiro83,Straumann84})
\begin{equation}
I_P(E;X,Y) \propto x^3 \exp(-x),
\end{equation}
where $x=E/kT(X,Y)$. \cite{Sha02} used similar expressions to study microlensing
in the optical continuum.
 Taking into account that the observed hard X-ray continuum has a
power-law type
spectral shape, we will also assume that the time-independent
intrinsic emissivity of the continuum is:
$$I(E,r)\sim E^{-\Gamma}\times r^{-\alpha},$$
where, according to the investigation of observed X-ray spectra,
$\Gamma$ and $\alpha$ are taken to be 1.5 and 2.5 (see
e.g. Dov\v ciak et al. 2004).
For the Fe K$\alpha$ emission in this case we used the same
calculation as in Popovi\'c et al. (2003a,b).

We should note here, that disk may be considered to be composed of
a number of distinct parts with different physical conditions (e.g.
radiation pressure dominant part, matter pressure dominant part, etc. see e.g. Shakura \&
Sunyaev 1973). Consequently, in general one can expect that the disk can be
described by different emissivity laws in different parts (e.g. the
black-body law may be applied in outer part of the disk). Taking
into account a huge number of parameters which should be considered in
the case of microlensed disk (see the next section), we will
consider only one emissivity law for whole disk.

 The total observed flux for the
continuum and the Fe K$\alpha$ line is given as
\begin{equation}
F(E)=\int_{\rm image} [I_C(X,Y;E)+I_L(X,Y;E)]d\Omega ,
\end{equation}
where $d\Omega$ is the solid angle subtended by the disk in the
observer's sky and the integral extends over the whole emitting
region.

As one can see from Eq. (7) the total observed flux is a sum of the
continuum and the line fluxes, consequently, the amplification in the
continuum and in the Fe K$\alpha$ line can be considered separately as two
independent components. On the other hand, the amplifications will depend
on the sizes and geometry of the continuum and line emitting
regions. In further text we will consider amplifications in the line and
in the continuum separately.

We would like to point out here that the aim of the paper is not to
create the perfect accretion disk
model (taking into account different  effects that can be present as
e.g. opacity of the disk, spots in the disk etc.),  but only to illustrate
the influence of microlensing on the continuum and the
Fe K$\alpha$ line amplification and demonstrate that this phenomenon could
essentially change
general conclusions. Therefore, we will use the three emissivity laws of
the disk and the very important effect of strong gravitation
(beaming and light-bending in Schwarzschild  and Kerr metrics).

\subsection{Disk and microlens parameters}

To apply the model one needs to define a  number of parameters that
describe the emission region and the deflector. In principle, we
should find constraints for the: size of the disk emission region,
disk inclination angle, mass of the black hole, accretion rate, relative
amplification, the constant $\beta$ (Chartas et al. 2002, Popovi\'c et al.
2003),
orientation of the  caustic with respect to the
rotation axis, direction of the caustic crossing and microlens
mass. In the following subsections we choose  and discuss the parameters
used in the calculations.

\subsubsection{Accretion disk parameters}

For the disk inclination we adopt the averaged  values given by
\cite{Nan97} from the study of the Fe K$\alpha$ line profiles of 18
Seyfert 1 galaxies: $i=35^\circ$.  The inner  radius, $R_{in}$, can not be
smaller than the radius of the marginally stable orbit,
$R_{ms}$, that corresponds to $R_{ms}=6R_g$ (gravitational
radii, $R_g=GM/c^2$, where $G$ is gravitational constant, $M$ is
the mass of central black hole, and $c$ is the velocity of light) in  the
Schwarzschild metric
and to $R_{ms}=1.23R_g$ in the case of the Kerr metric with
angular momentum parameter $a=0.998$. To select  the outer radius,
$R_{out}$, we take into account previous investigations of the X-ray
variability that support very compact X-ray  emitting disks. In
particular,  \cite{Osh02} from the observed variation in the lensed
blazar PKS 1830-211 infer a size of the X-ray  continuum emission region
of $\sim 3\times 10^{14}\rm cm$, that is in agreement with estimation
for QSO 2237+03050 given by \cite{Dai03}. So,
considering a range of black hole masses of  $10^7-10^9\ \rm M_\odot$ we
can conclude that the X-ray emission is coming from a compact
region of the order of 10 to 100 $R_g$. This  range of sizes is also
acceptable for the Fe K$\alpha$ emission region (see e.g.
\cite{Nan97,Nan99}).

To explore the suitability of the various  hypothesis explaining the
lack of adequate response of the X-ray continuum to  the microlensing
events detected in the Fe K$\alpha$ line (see \S 1),  we are going to consider
several combinations of disk sizes for the emitters of both the continuum
and the line: (i) the inner and outer radii of both emission regions are
the same, $R_{in}=R_{ms}$
and $R_{out}=20\ R_g$; (ii) the inner radius  is the same,
$R_{in}=R_{ms}$, but the outer radius of the  X-ray continuum disk is
smaller, $R_{out}=20\ R_g$, than the radius of  the line emission disk,
$R_{out}=80\ R_g$; (iii) the
continuum emission disk has radii $R_{in}=R_{ms}$,  $R_{out}=20\ R_g$ and
the line emission disk $R_{in}=20\ R_g$
and $R_{out}=80\ R_g$ (the continuum emission takes  place in an inner part of disk
surrounded by an annulus of Fe K$\alpha$ emission);  (iv) the continuum
emission disk has radii $R_{in}=20R_{g}$, $R_{out}=80\  R_g$ and the line
emission disk $R_{in}= R_{ms}$
and $R_{out}=20\ R_g$ (the Fe K$\alpha$ emission is  located in the inner
disk and the continuum emission in the outer annulus).

 We adopt the central object mass and accretion rate from \cite{BZ02}. We
 assume a black hole of mass $M_8=10^8 M_\odot$ and accretion rate
 $\mathop m\limits^.= 0.4$ in Eddington units $({\mathop {m}\limits^{.}}
= {\frac{{1.578{\mathop {M}\limits^{.}} _{26} }}{{3.88M_{8}}}
 })$. We will use this value in order to determine the effective
 temperature distribution. These values are in agreement with
 \cite{Wang03} where it was found that the majority of  QSOs have black
hole  masses in the range of $10^8-10^9\ M_\odot$, and accretion rates
ranging from 0.01 to
1 in units of the Eddington accretion rate.

It is difficult to discuss the validity of different emissivity laws for
demonstation of the X-ray emission (in the line as well as in the continuum),
but sometimes, as for example in the case of black-body emissivity law, the
emissivity at X-ray wavelengths can be extremely small compared with,
for example, optical wavelengths, and X-ray photons are emitted from
a quite small region.
In Fig. 1 (bottom), we presented the continuum shapes for different
emisivity laws used in the calculation (maximum of each is normalized to one).
The shapes of the continuum were calculated for different
dimensions of the  disk. As one can see from Fig. 1 (bottom), the shape of the
continuum strongly depends not only on emissivity law, but also on disk
dimensions.

\subsubsection{Microlens model and  parameters}

Different types of caustics can be used to explain the
observed microlensing events in quasars. Moreover, for the exact event one can model the
caustic shape to obtain different parameters (see e.g. Abajas et al.
2004, Kochanek 2004
for the case of Q2237+0305). In order to apply an appropriate microlens
model,  first  we will consider a standard microlensing
magnification pattern (Figure 2, left)
for the
Q2237+0305A image
with 16 Einstein ring radii (ERR) on a side and $\kappa=0.36$,
$\lambda=0.40$ and $\kappa_c=0$. The  mass of microlens is taken to be
1$M_\odot$. The simulation was made employing ray-shooting techniques that
send rays
 from the observer through the lens to the source plane (Kayser et al.
 1986; Schneider \& Weiss 1987; Wambsganss et al 1990a,b). We assume a
flat cosmological model with $\Omega
=0.3$  and $H_{o}= 70\ \rm km\ s^{-1} Mpc^{-1}$.

In Figure 2 we presented a comparison
between the projected magnification map in the source
plane and an accretion disk with a size of 1000 R$_g$ (presented as
a circle in Figure 2, right).
Taking into account the small dimensions of the X-ray emission region
(several 10s R$_g$) the approximation of a straight fold caustic can be
assumed for this pattern. We explored the
general behavior of the total continuum and Fe K$\alpha$ flux amplification due to microlensing.
First we used the straight fold caustic approximation. (see
Eqs. (5)-(8) in \cite{Pop03}). We also considered an example assuming
a caustic magnification pattern for the Q2237+0305A image produced by a
population of low mass microlens in \S 3.3.

 In Table 1 we give the projected Einstein Ring Radii (ERR) for the
 lensed QSOs where amplification of the Fe K$\alpha$ line  (MG
J0414+0534,
 \cite{Chart02a};  QSO H1413+117, \cite{Osh01,Chart04} and QSO
 2237+0305, \cite{Dai03}) has been observed. The ERRs (expressed in
 gravitational radii) are computed for different deflector masses and
 for a black hole mass of $10^8\ \rm M_\odot$. We found that even deflectors
 with small mass have ERR with sizes from several tens to
several hundreds of gravitational radii.
 To obtain a qualitative understanding of the influence of the microlens
mass on our results
we consider microlens masses that correspond to ERR values equal to 50
R$_g$ (see Figs. 3-7).
 Qualitatively  the shape of the flux
amplification
will not be
changed if we consider massive deflectors (e.g. ERR=2000 $R_g$, see
Fig. 8). Even if we apply  simple assumption of the straight-fold
caustic,
we should define the caustic parameters A$_0$ and $\beta$. These values
can be considered to be different for different microlensing events.
Higher values of A$_0$ and $\beta$ will cause higher amplification. Here
we would like to demonstrate correlation between the line
and flux amplification due to microlensing and we will adopt values considered
in more details by Chartas et al. (2002)  $A_0$=1 and $\beta$=1 (Witt et al. 1993). Here we
considered three directions of caustic crossing; parallel (Y=0) and
perpendicular (X=0)
to the rotation axis as well as in a direction inclined 45$^\circ$ with
respect to the rotation axis (X=Y).

\section{Results and Discussion}

\subsection{Continuum and line  profile variability}

 In Figures 3 and 4 we present the variations of the total X-ray emission
 spectra (continuum + Fe K$\alpha$ line) during a straight fold
 caustic crossing ($A_0$=1, $\beta$=1 and ERR=50 $R_g$).  The radial
 dependence of the emissivity is related to the black body radiation
 law (see \S 2). In Figure 3 the sizes of the continuum and line emission
 regions are the same, $R_{\rm inn}=R_{\rm ms}$ and $R_{\rm
 out}=20\ R_g$. In Figure 4, the Fe K$\alpha$ line emitting disk is
 larger: $R_{\rm out}=80\ R_g$. We consider both metrics,
 Schwarzschild and Kerr. We simulate the caustic crossing perpendicular to
 (first and second rows) and along (third and fourth rows) the
 rotation axis in both directions; $\kappa=\pm 1$, respectively.

\subsection{Integrated flux variability}

 In Figures 5-8 we present the variation of the integrated flux
(normalized to the integrated flux in the absence of microlensing) for
 both, the X-ray continuum and the Fe K$\alpha$ line during straight fold
 caustic crossings (considering that the amplification outside the
caustic $A_0$=1, as well as  $\beta$=1 and ERR=50 $R_g$, see Figure
captions).
 We considered all three emissivity laws (see \S 2.1)
 Figures 5a,b correspond to case (ii) of \S 2.2.1, Figures 5c,d and 7 (left) to  case
(iii) and Figures 6 and 7 (right) to
 case  (iv). (Case (i) is  the same for line and
continuum; it corresponds to the continuum variation in Figures 5a,b). {  In Fig. 7,
we present  cases iii) and iv) where a power-law emission is taken into
account.} In Figure 8 a very favorable case of high ERR and inclination is considered for case
(iii) of  \S2.2.1. Notice that  FeK$\alpha $ microlensing events were
observed in BAL QSOs which may have highly inclined accretion disks; $i\approx
75^\circ$.

{ As one can see from Figure 8 (as it also was shown in Popovi\'c et
al. 2003a,b) the amplified component is mainly very narrow in
comparison with the undeformed line. This result is in agreement with
the observations of Chartas et al. (2002, 2004) and Dai (2003) and
supports the conclusions of these authors that
enhancement of the Fe K$\alpha$ line observed in only one of the
images of a lensed quasar was caused by microlensing.
}

 Some interesting results can be inferred from our exploratory
 work: (a) when both the line and continuum disk profiles have the same
inner radius, the
 differences in outer radius cannot cause significant differences in the
total line and the continuum flux variation (see Figs. 5a,b) for the
considered emission laws,
 (b) when we separate the emission considering an inner
 disk contributing to the continuum and an outer annulus contributing to
the Fe K$\alpha$ line (or vice-versa) we
 found significant differences between the continuum and the line
 amplification (see Figures. 5c,d, 6 -- 8),  (c) the results are
qualitatively similar for {all considered emissivity laws} (see
Figs. 5-8),
 (d) interchanging between
Schwarzschild and Kerr metrics induces only slight
 differences in normalized flux amplification (see Figs. 5-7).

 Figures  5-8 were intended to explore the two scenarios suggested by
\cite{Dai03} and \cite{Chart02a} to explain the non
observed associated enhancement of the X-ray continuum in objects with
a microlensed
Fe K$\alpha$ line. This behavior can be expected
in  case (iii) when the microlens crosses the outer part of the disk
(Figures 5c,d and the right panel of Fig. 7) and in case (iv) when the
microlens
crosses the inner part of the  disk (see Fig. 6 and the left
panel of Fig. 7). However, in none
of the Figures does the continuum remain strictly constant
during a complete Fe K$\alpha$  microlensing event.  In the most favorable
case (the inner Fe K$\alpha$ disk plus an outer
continuum annulus; Fig. 6 and 7) we achieve a significant
and relatively quick change of the Fe K$\alpha$ emission while the
continuum experiences only a slow increase.  This behavior could
well approximate a non-varying continuum
but only if we consider observations in a temporal  window
that fall on the peak of the microlensing event in the Fe K$\alpha$ line.
 In this case the continuum of the  microlensed image
experiences an (slowly changing or
almost constant) amplification with respect to the  continua of the other
images, but practically it is indistinguishable from the amplifications
due to global macrolensing.

\subsection{Microlensing by a caustic magnification pattern: An example
for Q2237+0305A image}

Here we consider a situation where a low mass population of
microlenses (smaller than one solar mass)  can form pattern structures (see Table 1)
which are comparable with the
size of the X-ray accretion disk.
Moreover, the black hole mass of
the lensed quasar may be of the order of $10^{9-10}M_\odot$, taking that
$R_g\sim M_{BH}$, the pattern structure of low mass microlenses are
comparable with a
X-ray disk size of several dozens R$_g$. Therefore, here we consider that the
black hole mass of the
lensed quasar is $10^{9}M_\odot$.

For modeling of the caustic magnification pattern for image Q2237+0305A
we used the same values for the convergence and external shear as
 presented in Figure 2, but for a low mass population, taking that the
mass of the deflectors are
 randomly distributed in an interval ranging from 0.1 $M_\odot$ to 0.6
$M_\odot$,
with a mean value of $<m>=0.35\ M_\odot$.
Also, the positions of the lenses were distributed randomly in a rectangular
 region in the lens plane, significantly larger than the considered region
 in the source plane. Now, $1ERR$ projected in the source plane
corresponds to

$$
 ERR(M)= ERR(M_{\odot}) \sqrt{{\frac{M}{M_{\odot}}} \frac{M_{8}}{M_{BH}}},
$$
where $ERR(M_{\odot})=0.054$pc is the projected ERR for a solar mass
deflector, $M$ is the mean mass of the deflectors and $M_{BH}$ is the
black-hole mass. Taking the mean deflector mass as $<m>=0.35\ M_\odot$
and $R_g=9.547.10^{-6} M_{BH}/M_{8}$ pc,
we modeled a caustic magnification pattern of 1ERR$\times$2ERR, that
corresponds to 334.63R$_g \times$669.26$R_g$ in the source plane for a
black hole mass of
$M_{BH}=10^{9}M_\odot$ (Fig. 9). For numerical reasons, the  microlens
magnification  map  is given in pixels, 1000$\times$2000
(1pix=0.33463R$_g$ in source plane).  As one can see from Figure 9, the
microlensing pattern structures are comparable with a compact X-ray
accretion disk.

In our previous modeling based on  the straight fold caustic
approximation the lack of a correlation between the continuum and Fe
K$\alpha$ line is expected only if  the line and X-ray continuum
region are separated. Recent investigations of the Fe K$\alpha$
line profile from active galaxies show that the line should be emitted
from the innermost part of the accretion disk. In particular,
Ballantyne \& Fabian (2005) found that in BLRG 4C+74.26 the outer radius
of the relativistic iron line should be within 10 R$_g$.
 Consequently, here we will assume that the
 Fe K$\alpha$ line is formed in the innermost part of the disk ($R_{\rm
inn}=R_{\rm ms}$; $R_{\rm
out}=20$ R$_{g}$) and that  the continuum (emitted in the
energy range between 0.1 keV and 10 keV) is mainly
originated from a larger region ($R_{\rm
inn}=20$ R$_{g}$; $R_{\rm
out}=100$ R$_{g}$).\footnote{Note here, taking the continuum disk
size from $R_{\rm
inn}=20$ R$_{g}$ to $R_{\rm
out}=100$ R$_{g}$, we
neglected the contribution of the innermost part emission (from R$_{ms}$
to 20 R$_g$)
to  the total continuum flux only in the energy interval from 0.1 keV to
10
keV. It does not mean that there is no the continuum emission.}
On the other hand, from the straight fold caustic modeling we
conclude that
the correlation between the total line and continuum flux due to microlensing
is not very different for different emissivity laws. Consequently, here we
used the black-body emissivity law. A disk (Schwarzschild metric) with
an inclination of 35$^{\circ}$ is considered.

To explore the line and X-ray continuum variation we moved the disk center
along the microlensing map as it is shown in Figure 9 (from left to the right
corresponding 0 to 2000 pixels). In Figure 10 we present the
corresponding total line and X-ray continuum flux variation. As one can
see from  Figure 10,
there is a global correlation  between the total line and continuum flux
during the complete path.
However, the total continuum flux  variation  is smooth and has a
monotonic  change,
 while the total line flux  varies very strongly and randomly.

In fact, during some portion of the microlensing of the emission regions by the magnification pattern,
we found the total
Fe  K$\alpha$ line flux changes, while the continuum flux remains
nearly constant (e.g. the position of the disk center between 1000 and
1200 pixels).
This  and the shapes of the line and continuum total flux amplification
indicate that   the observed microlensing amplification of the Fe
K$\alpha$ in three lensed quasars may be explained
if the  line is originated in the innermost part of the
disk and the X-ray continuum in a larger region.  Also, it
seems that the contribution of the continuum emitted from the innermost part of the
disk (within  10 R$_g$)  to the total continuum (in the energy
interval
from 0.1 to 10 keV)  flux is not significant. Further observations are
needed to
provide more data which might be compared with our theoretical
results.

\subsection{Wavelength dependence of the X-ray continuum amplification}

 The influence of gravitational microlensing on the
spectra of lensed QSOs was discussed in several papers
(see Popovi\'c  and Chartas 2005, and references therein). Mainly, the
color index was calculated as an indicator of the microlensing (see
e.g. Wambsganss \& Paczynski 1991, Wyithe et al. 2000) as well as
amplified flux behaviors (see Yonehara et al. 1998, Yonehara et al.
1999, Takahashi et al. 2001)
of a disk,
but the exact shape of the amplification as a function of wavelength
(or energy) for a partly microlensed disk has not
been calculated. Here, taking into account that
 the emitters at different radii in the accretion disk have different
temperatures (see Fig. 1) and make different contributions to the
 observed continuum flux at a given wavelength, we calculated the amplification as a function of observed energies.
During a caustic crossing
microlensing effects would depend on the location
 of the emitters and, consequently, would induce a wavelength dependence
in the amplification. This dependence can be
 clearly appreciated in the spectra of Figures 3, 4 and 8. In Figures
11ab-12ab
we present the amplification as a function of
 the observed energies for an accretion disk with the
characteristics given in  \S 2.2.1, with inner radius
 $R_{in}=R_{ms}$ and outer radius $R_{out}=30\ R_{g}$, assuming caustic
crossing along the rotation axis (X). We have
 considered the black body (Fig. 11ab) and the modified black body
(Fig. 12ab) emissivity laws for both Schwarzschild and Kerr
metrics.

 As it can be seen in Figures 11-12 the amplification is different for different
observed  energies. The
 amplification is  higher for the hard X-ray continuum when the caustic
 crosses the central part of the disk (see
 Fig. 12).   Depending mainly on the caustic location and on the
emissivity law selected the difference of the
 amplification in the energy range studied by us can be significant
(e.g.  $\sim 20$\% for very small mass microlenses, ERR=50$R_g$, see
Figs. 11-12). This effect could induce a noticeable
 wavelength dependent variability of the X-ray continuum spectrum during a
microlensing event (of even a 30\%), providing a tool to study the
 innermost regions of accretion disks.

\section{Conclusions}

 We have developed a model of microlensing by a straight fold caustic of
 a standard accretion disk in order to discuss the observed enhancement of
 the Fe K$\alpha$ line in the absence of corresponding continuum
amplification found in
 three lensed QSOs. Here we summarize several interesting results inferred from our straight
fold caustic simulations.

 1 - As expected both the Fe K$\alpha$ and the continuum may experience
significant
amplification by a microlensing event (even for microlenses of very small mass).  Thus, the absence of
adequate continuum amplification in the observed Fe K$\alpha$ microlensed QSOs should be related to the structure
of the accretion disk and/or the geometry of the event.

 2 - Extending the outer radius of the distribution of Fe K$\alpha$
 emitters does not result in any significant changes in our results. This
is due to the radial dependence
of the emissivity as expressed in the standard
 accretion disk model
 \citep{Shakura73} that concentrates  the emission near the center of the
black hole making
negligible the
 contribution from the outer parts to the integrated flux. In principle we
 could consider other less steep emissivity laws that  make
 the outer parts of the disk more important, however previous studies
 (e.g. \cite{Nan99}) support the hypothesis of a strong emissivity
gradient.

  3 - Segregation of the emitters allows us to reproduce the Fe K$\alpha$
enhancement without equivalent continuum
 amplification if the continuum emission region lies interior  to the Fe
K$\alpha$ emission region or vice versa but
 only during limited time intervals. In fact, in none of the simulations
does the continuum remain constant during a complete
 Fe K$\alpha$ microlensing event. In the case of an inner Fe K$\alpha$
 disk plus an outer continuum annulus a significant change of the Fe
 K$\alpha$ emission is obtained while the continuum experiences only a
very shallow
 gradient. This behavior could well approximate
the detected non-varying continuum
for observations covering only the peak of the Fe
K$\alpha$ microlensing event.

4 - We have also studied a more realistic case of microlensing by  a
caustic magnification  pattern assuming a population of low mass
deflectors. In this case we can successfully reproduce the observed lack of
correlation between the X-ray continuum and Fe K$\alpha$ emission
amplification only if the line and continuum emission  regions are
separated.

An extreme case of
 segregation is that of the two component model for the continuum
suggested in several papers (see e.g. \citep{Fab03,Pag04} if the
contribution of
the disk component to the
X-ray continuum were weak enough.

  5 - We have studied the chromatic effects of microlensing in the X-ray
continuum and  find that the dependence with wavelength of the
  amplification can induce, even for small mass micro-deflectors
 (ERR=50$R_g$), chromatic variability of about 30\% in the observed energy
range (from 0.1 keV to 10 keV) during
a microlensing event.

Further observations of the lensed quasars in the X-ray are
needed to  confirm these results. In any case monitoring of
gravitational lenses may help
us to understand the physics  of the innermost part of the
relativistic accretion  disks.

\begin{acknowledgements}
This work is a part of the projects: P1196 "Astrophysical Spectroscopy of
Extragalactic Objects" supported by the Ministry of Science, Technologies
and Development of Serbia and P6/88 "Relativistic and Theoretical
Astrophysics" supported by the IAC. L\v CP was supported by Alexander
von Humboldt Foundation through the program for foreign scholars.
JAM is a {\it Ram\'on y Cajal Fellow} from the MCyT of Spain.
Also, we would like to thank the anonymous referee for very useful comments.
\end{acknowledgements}


\begin{figure}
\begin{center}
\plotone{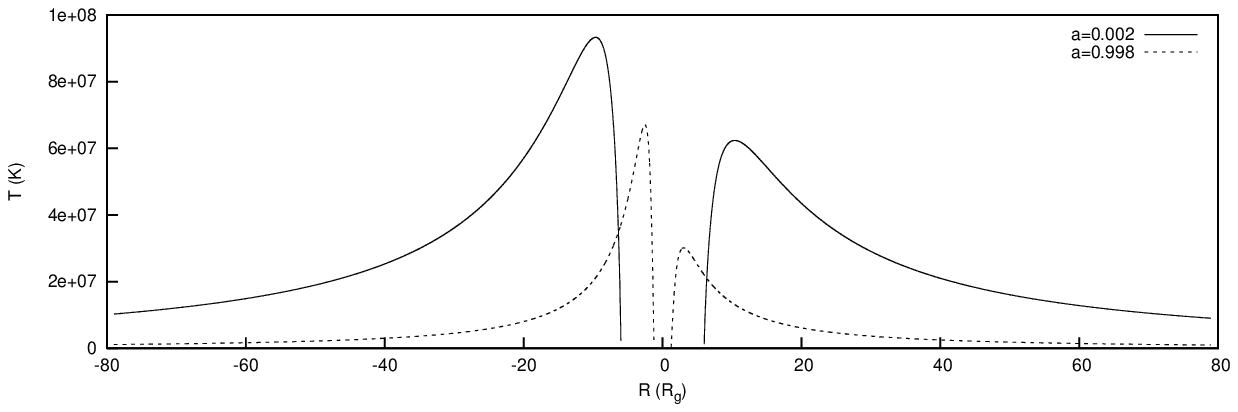}
\plottwo{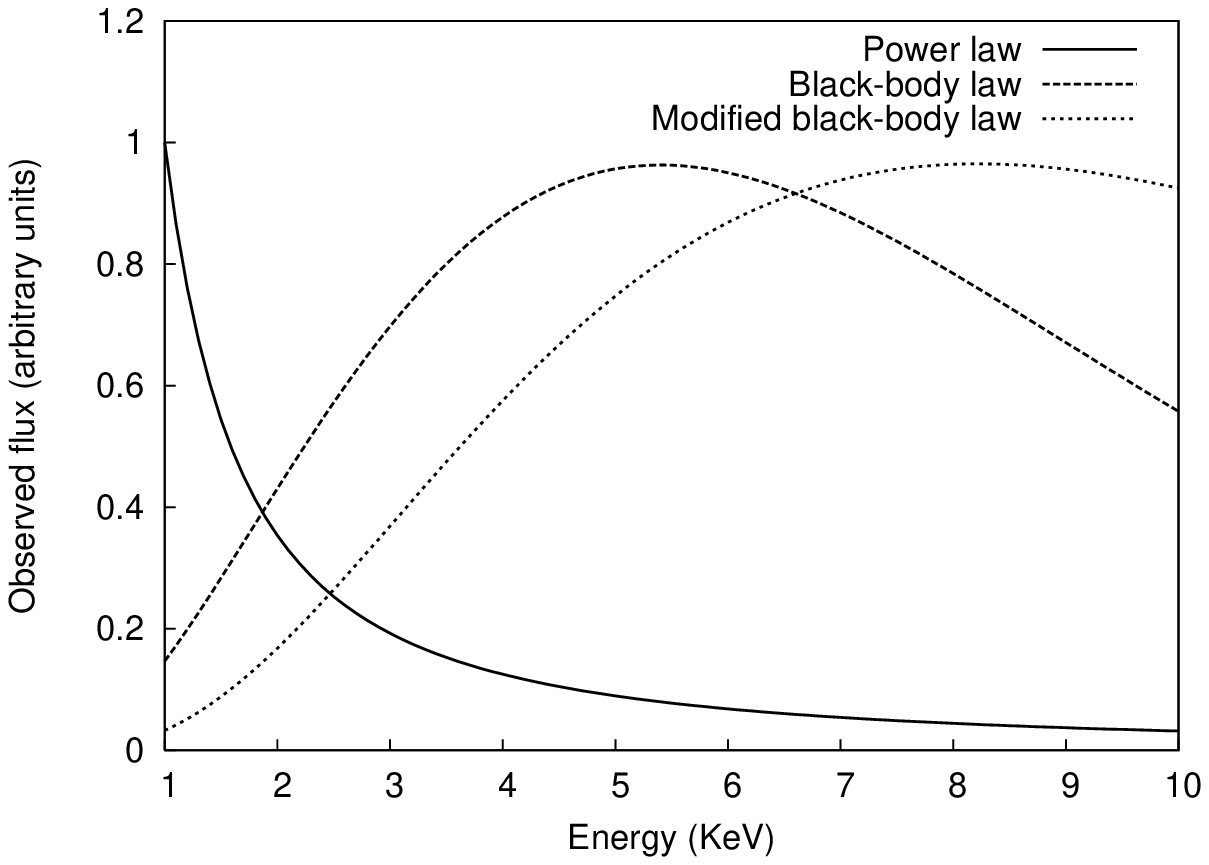}{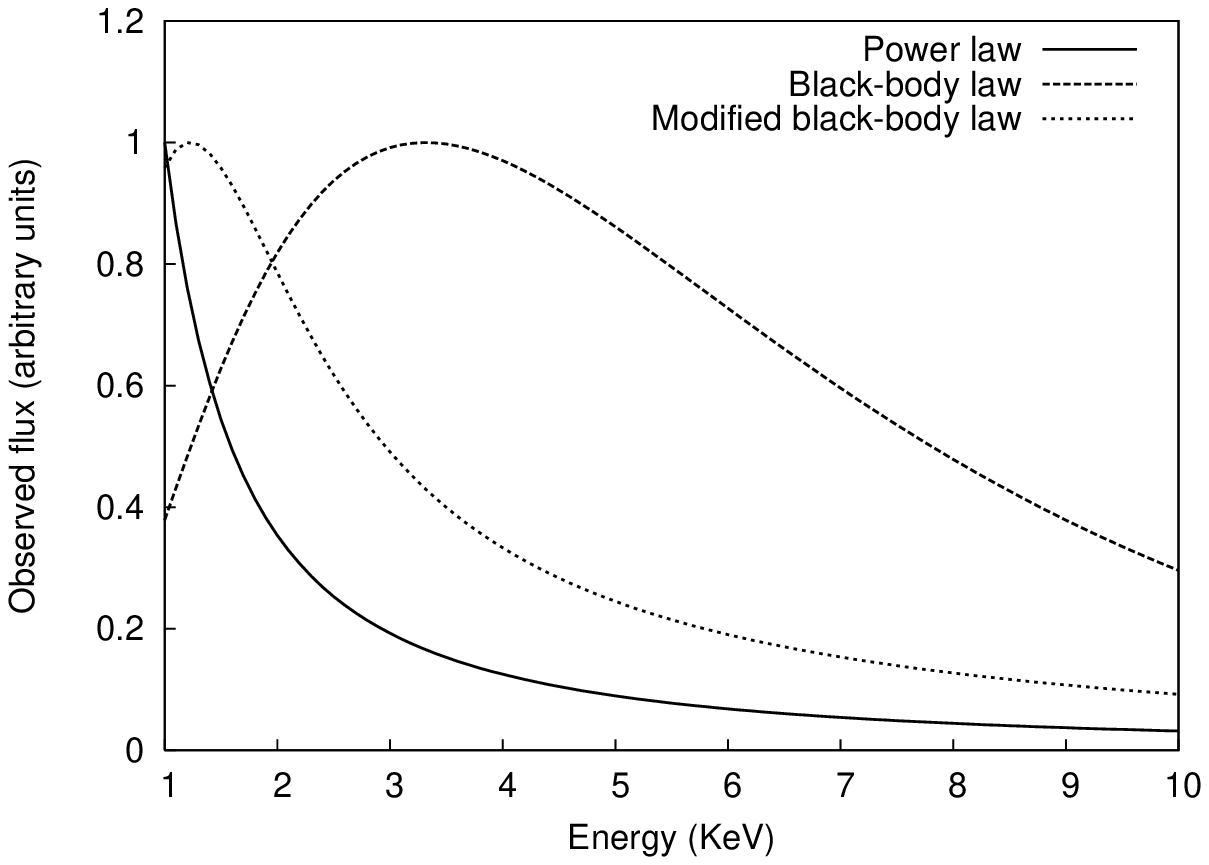}
\caption{Top:  The distribution of the temperature as a function of the
radius along the  direction of the disk rotation, given for two diferent
values of
angular momentum  $a$. Negative values of $R$ correspond to the approaching and positive values to the receiding side of the disk. Bottom: Shapes of the contunuum for considered three
emissivity laws (normalized to the maximal value) for an accretion disk
with outher radius of 20 R$_g$ (left) and 80 R$_g$ (right). The other
parameters of the disk are given in \S 2.2.1.}
\end{center}
\end{figure}
\clearpage

\begin{figure}
\begin{center}
\plottwo{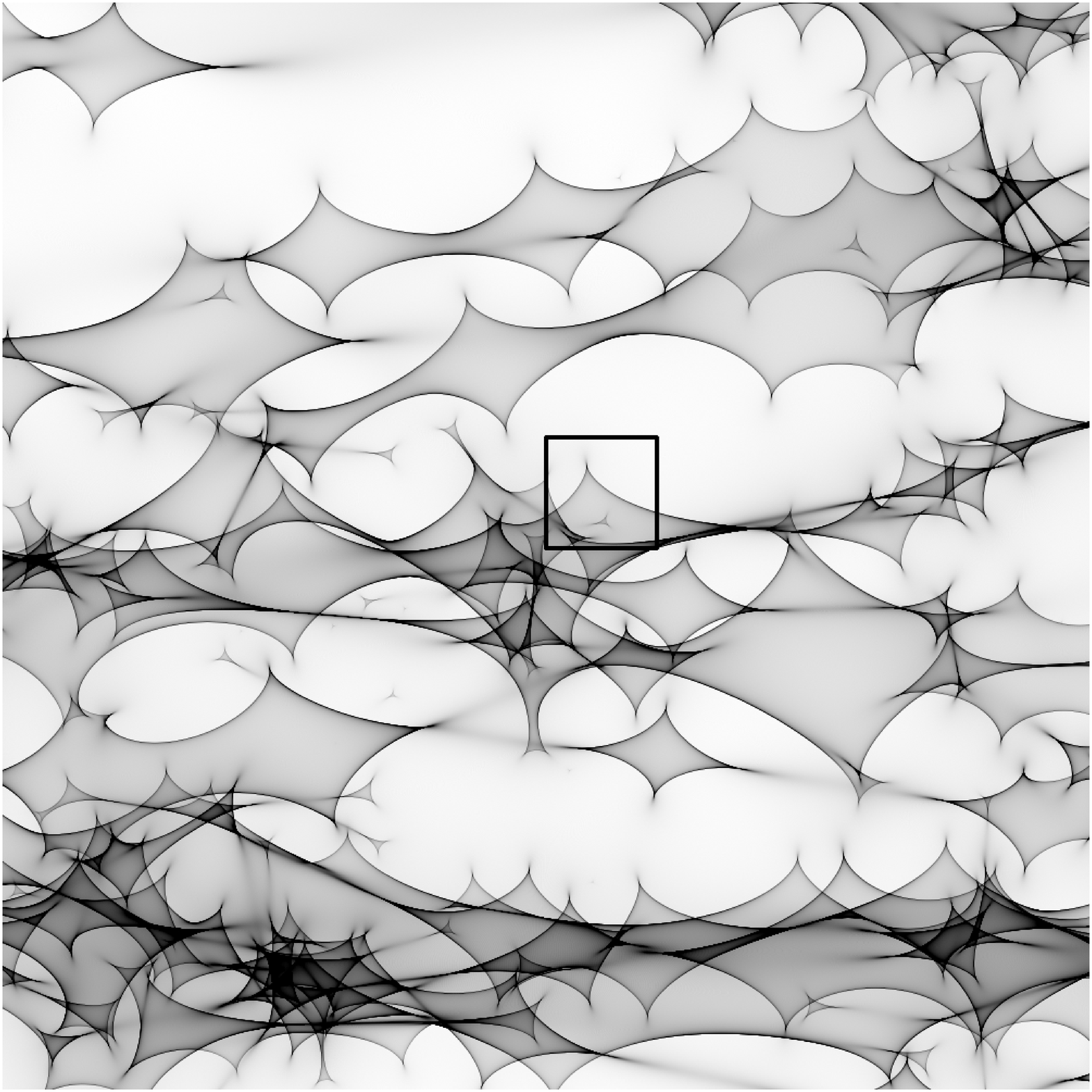}{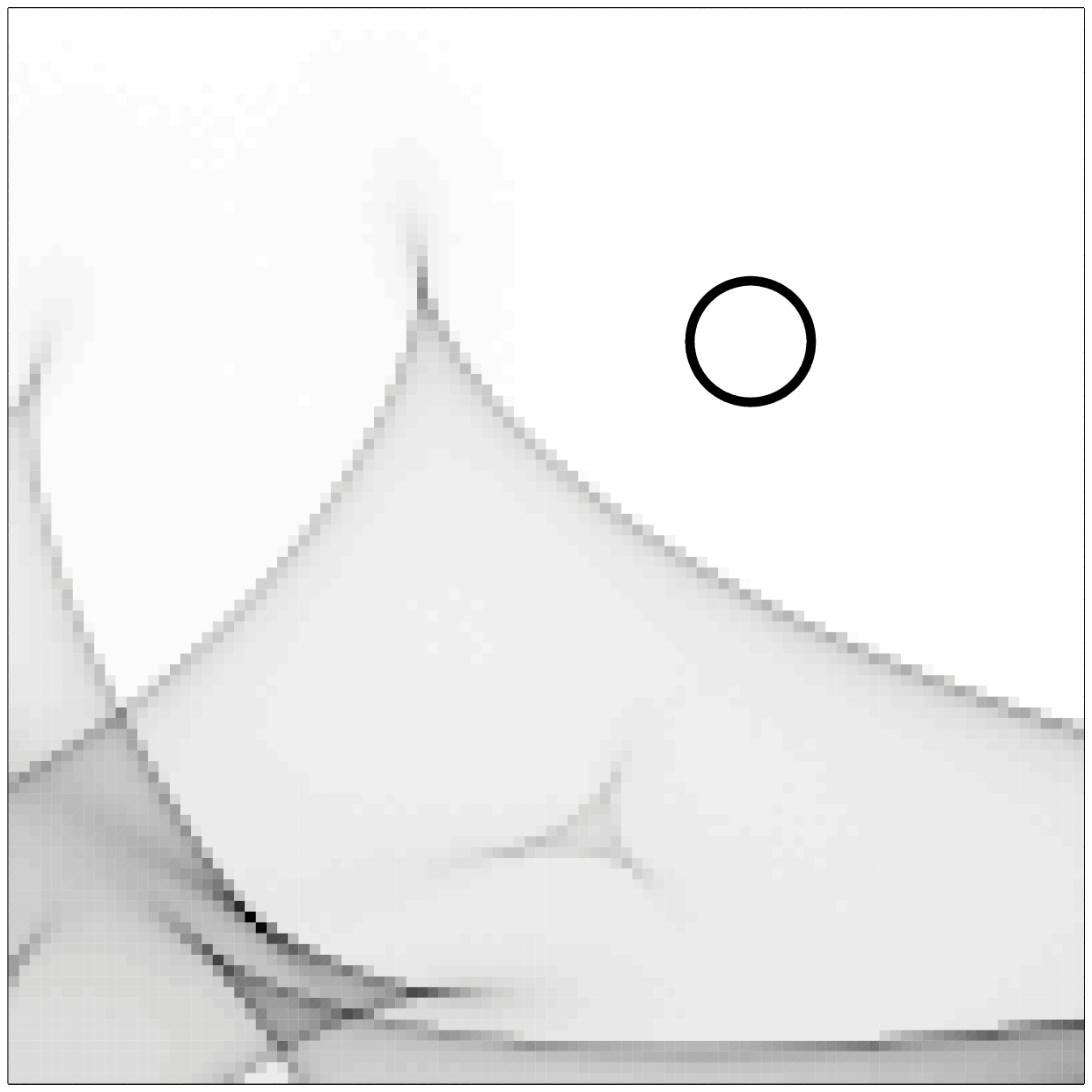}
\caption{{\it Left:} Microlensing map of QSO 2237+0305A image with 16 ERR
(177372 $R_g$) on a side (Abajas et al. 2004). {\it Right:} the
small part (square in Figure left) of the microlensing pattern,
compared to a face-on accretion disk. The
assumed outer radius of the disk is $R_{out}=1000\ R_g$.}
\end{center}
\end{figure}

\begin{figure}
\begin{center}
\plottwo{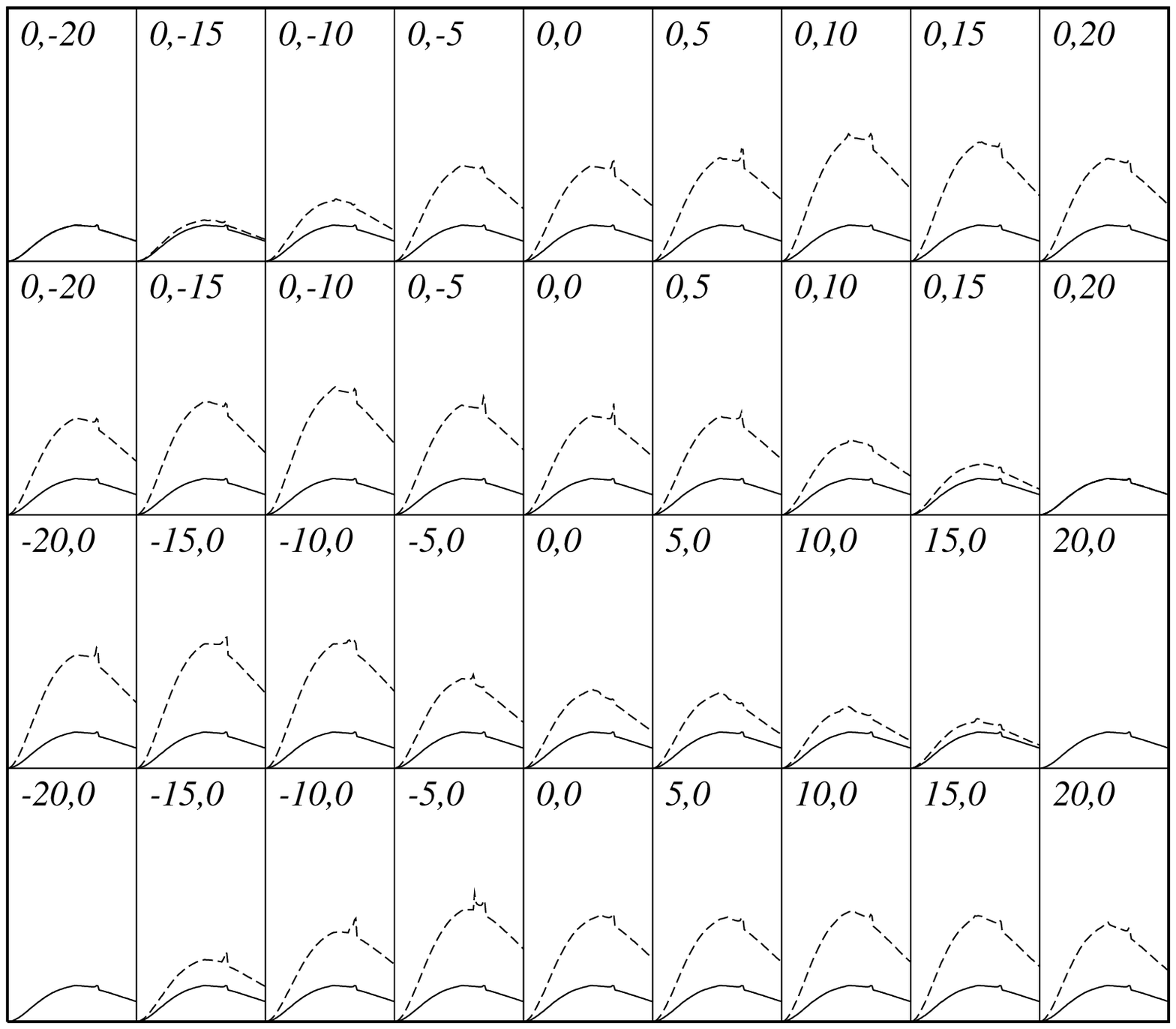}{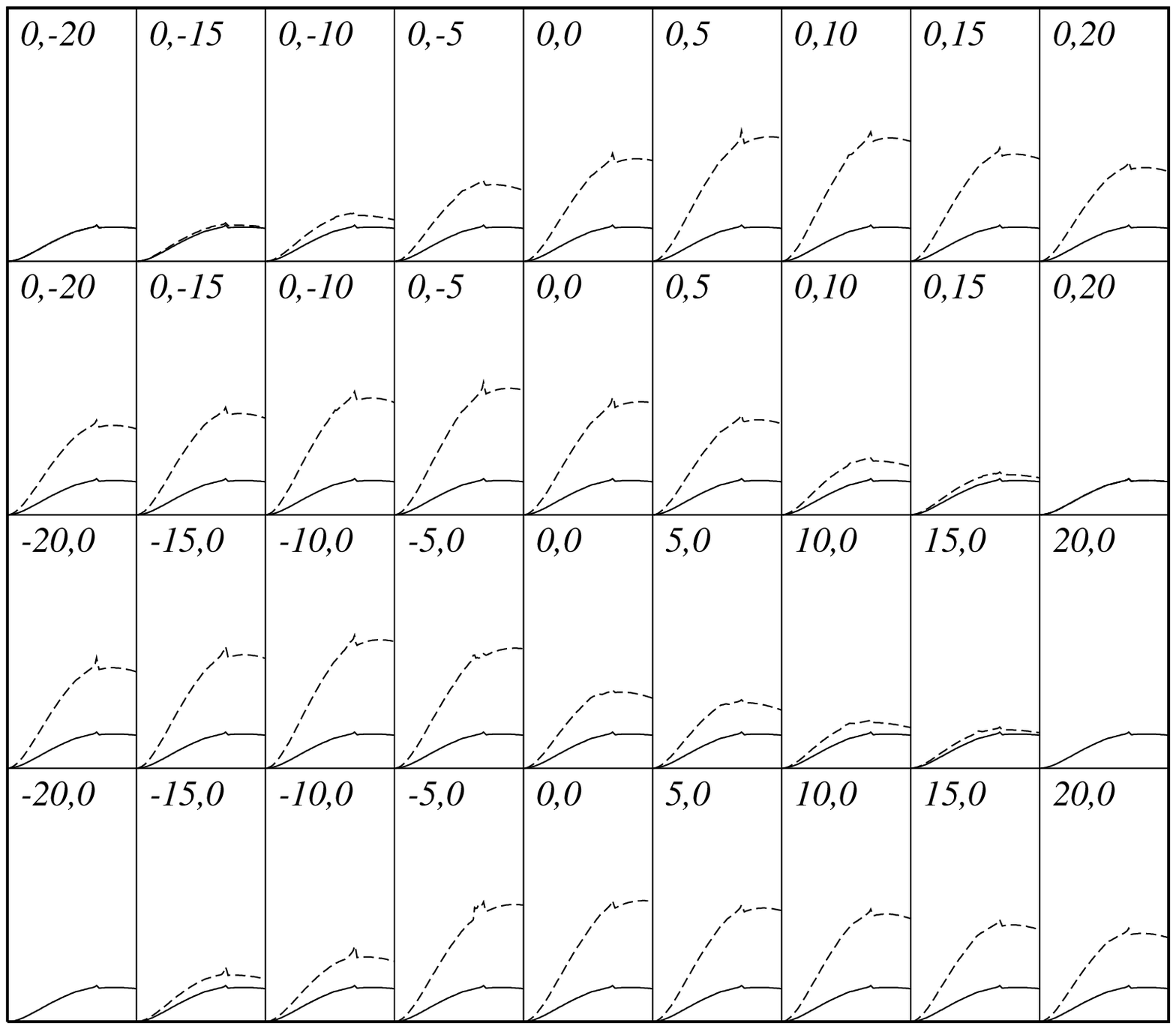}
\caption{Simulations of the behavior of the Fe K$\alpha$ line
and continuum variations due to microlensing by a caustic in the
case of Schwarzschild (left) and  Kerr (right) metrics. The
parameters of the caustic are: $A_0$=1, $\beta$=1 and ERR=50
R$_g$. In the first and second rows we present  the caustic crossing
perpendicular to the rotating axis for $\kappa=\pm1$, respectively.
 In the third and fourth rows we show the caustic
crossing along the rotation axis with $\kappa=\pm1$, respectively.
The radii of the continuum and the Fe K$\alpha$ line emission accretion
disks are the same: $R_{in}=\ R_{\rm ms}$ and $R_{out}=20\ R_g$ (where
$R_g=GM/c^2$).
The unperturbed and normalized emission correspond to solid and dashed
lines, respectively. The relative intensity ranges from 0 to 7
(y-axis)
and the energy interval from 0.1 to 10 keV (x-axis).}
\end{center}
\end{figure}

\begin{figure}
\begin{center}
\plottwo{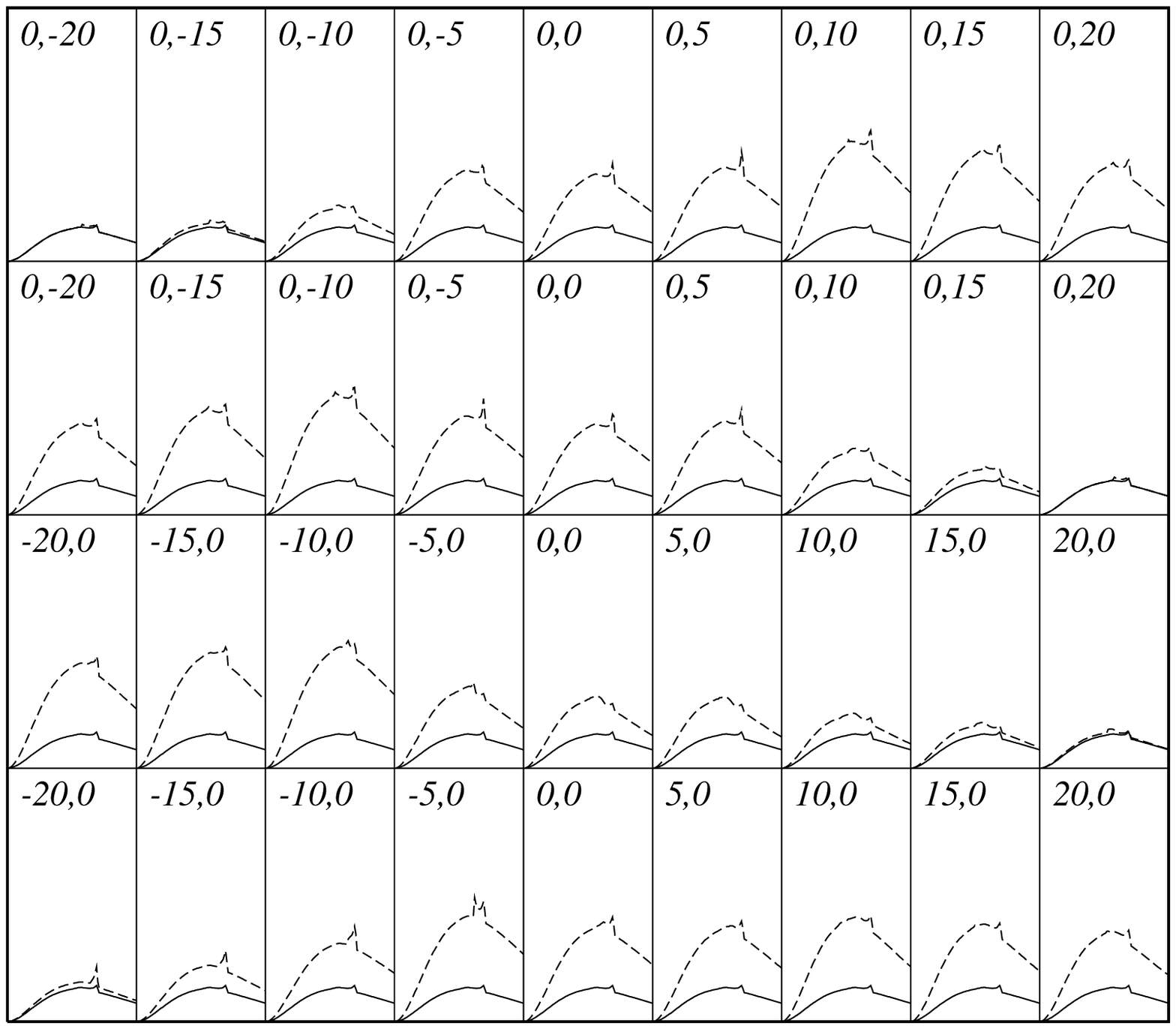}{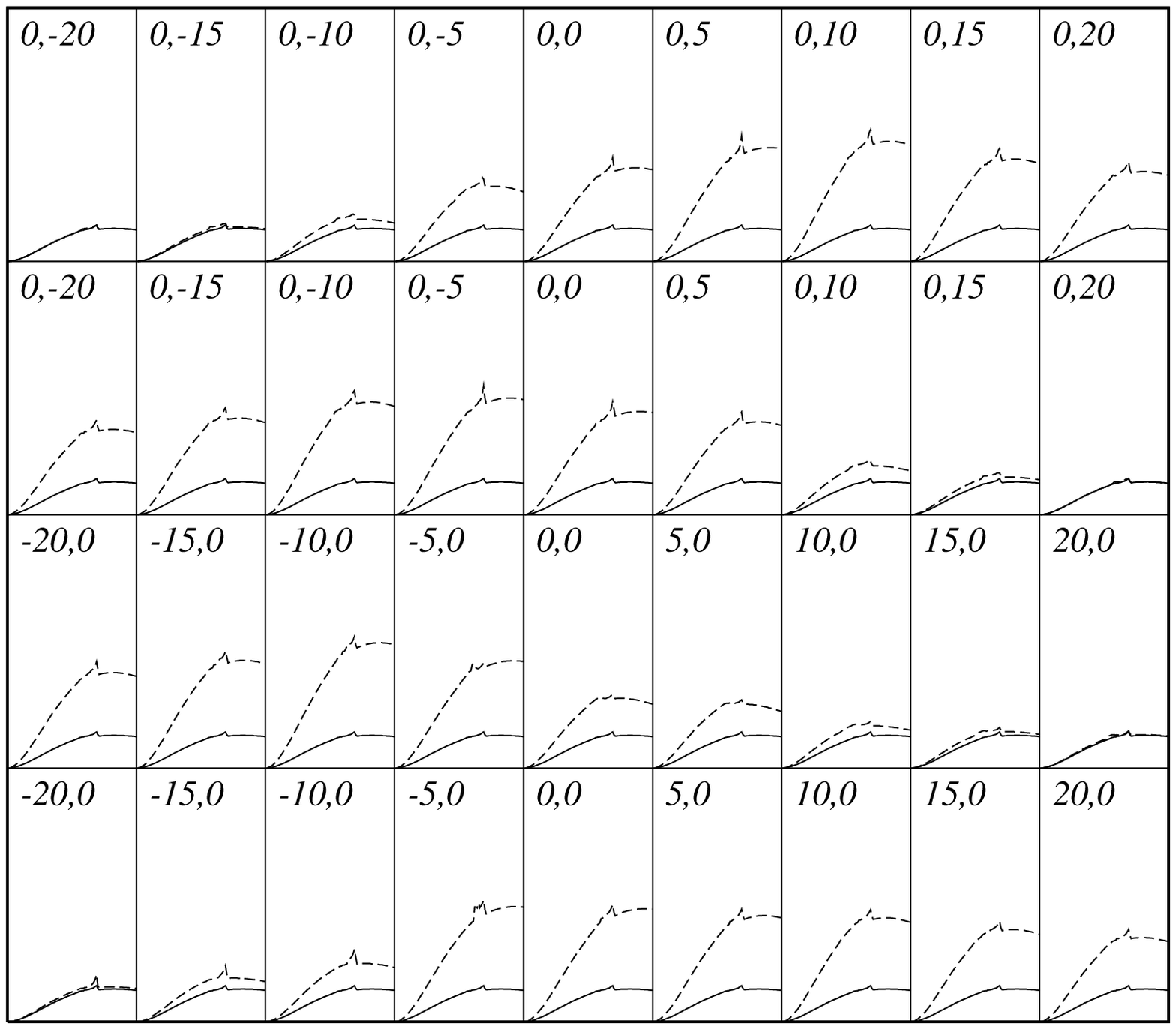}
\caption{As in Fig. 3, but
 $R_{out}=80\ R_g$ for the Fe
K$\alpha$ line.}
\end{center}
\end{figure}

\begin{figure}
\begin{center}
\plotone{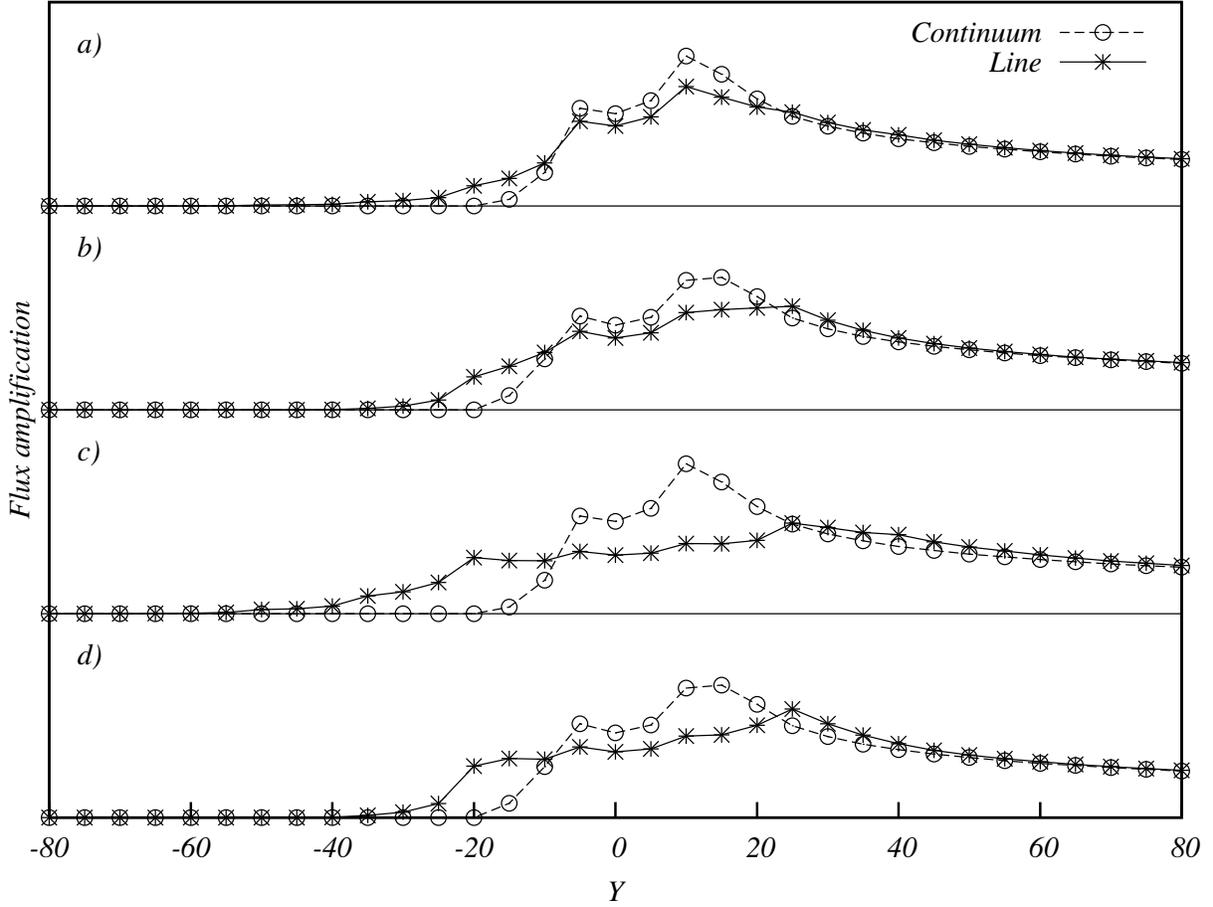}
 \caption{Variations of the Fe K$\alpha$ line and X-ray continuum flux due
to caustic crossing (ERR=$50\ R_g$)
 perpendicular to the rotation axis (X=0) for different positions of the
caustic in the Y axis
(Schwarzschild
metric). Panels correspond to: (a) black body law and case
(ii) of \S2.2.1,
(b) modified black body law and case
(ii), (c) black body law
and case (iii),
(d) modified black body law and case (iii).
The flux axis ranges from 1 to 1.7.}
\end{center}
\end{figure}

\begin{figure}
\begin{center}
\plottwo{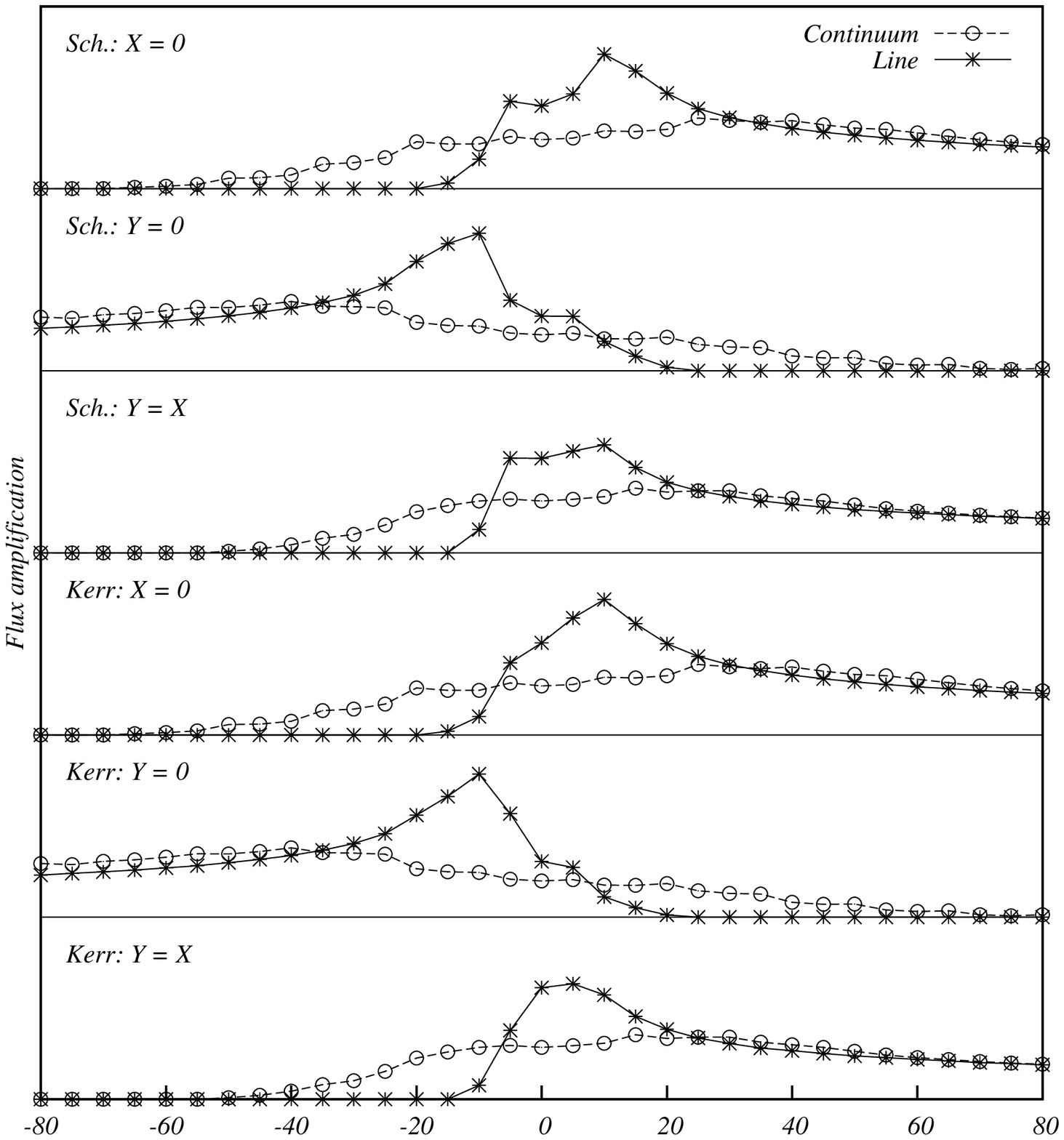}{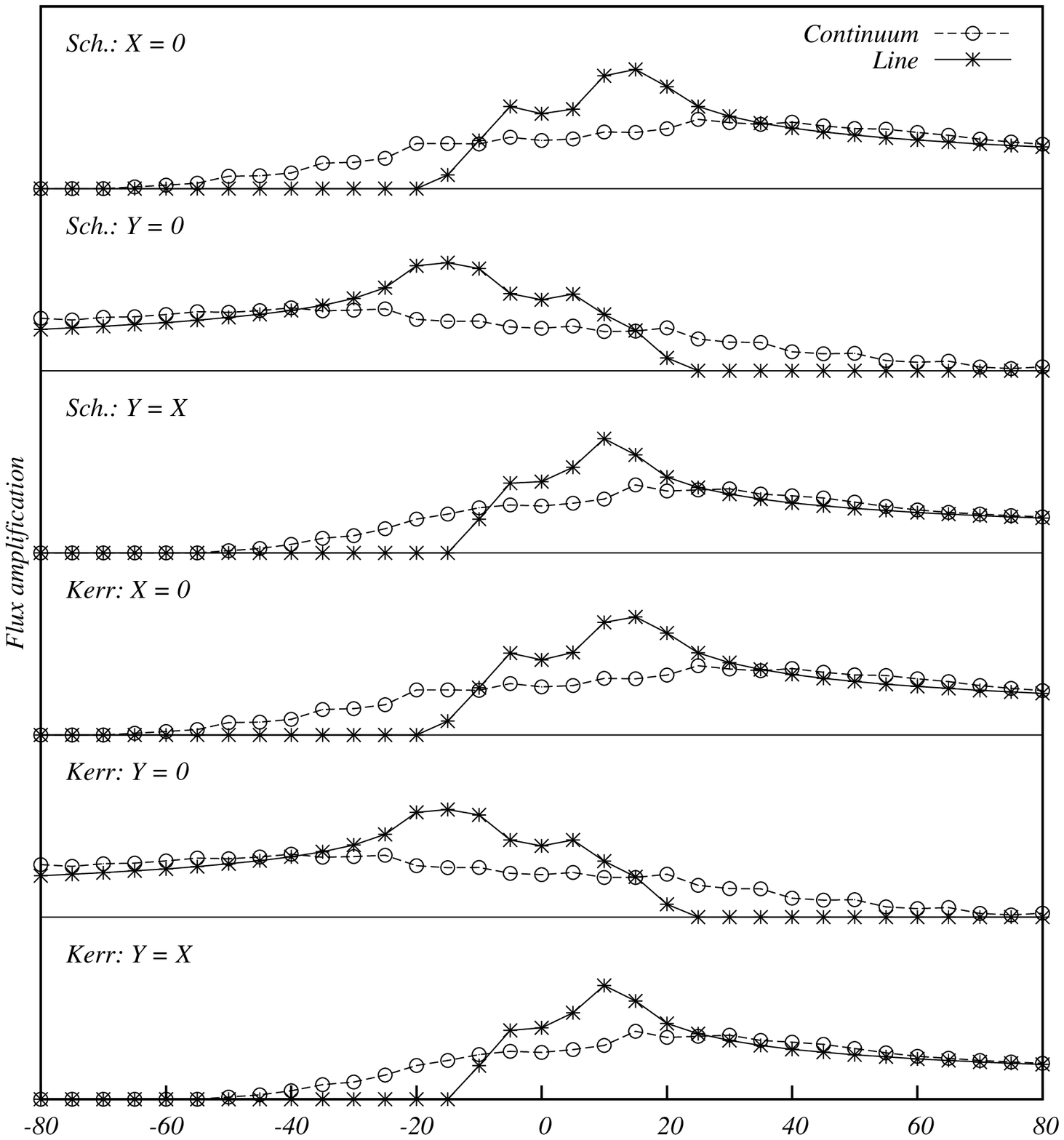}
\caption{Variations of the Fe K$\alpha$ line and
 X-ray continuum flux during a caustic crossing
(ERR=$50\ R_g$) for the case (iv) of \S 2.2.1. We consider three different directions of caustic crossing (X=0, Y=0 and X=Y; see \S 2.2.2). Left and right panels correspond
to the black body
and  modified black body laws, respectively. From top to bottom, the three first
panels (left and right)
correspond to Schwarzschild and the last three to Kerr metrics.
The flux axis ranges from 1 to 1.7.}
\end{center}
\end{figure}

\begin{figure}
\begin{center}
\plottwo{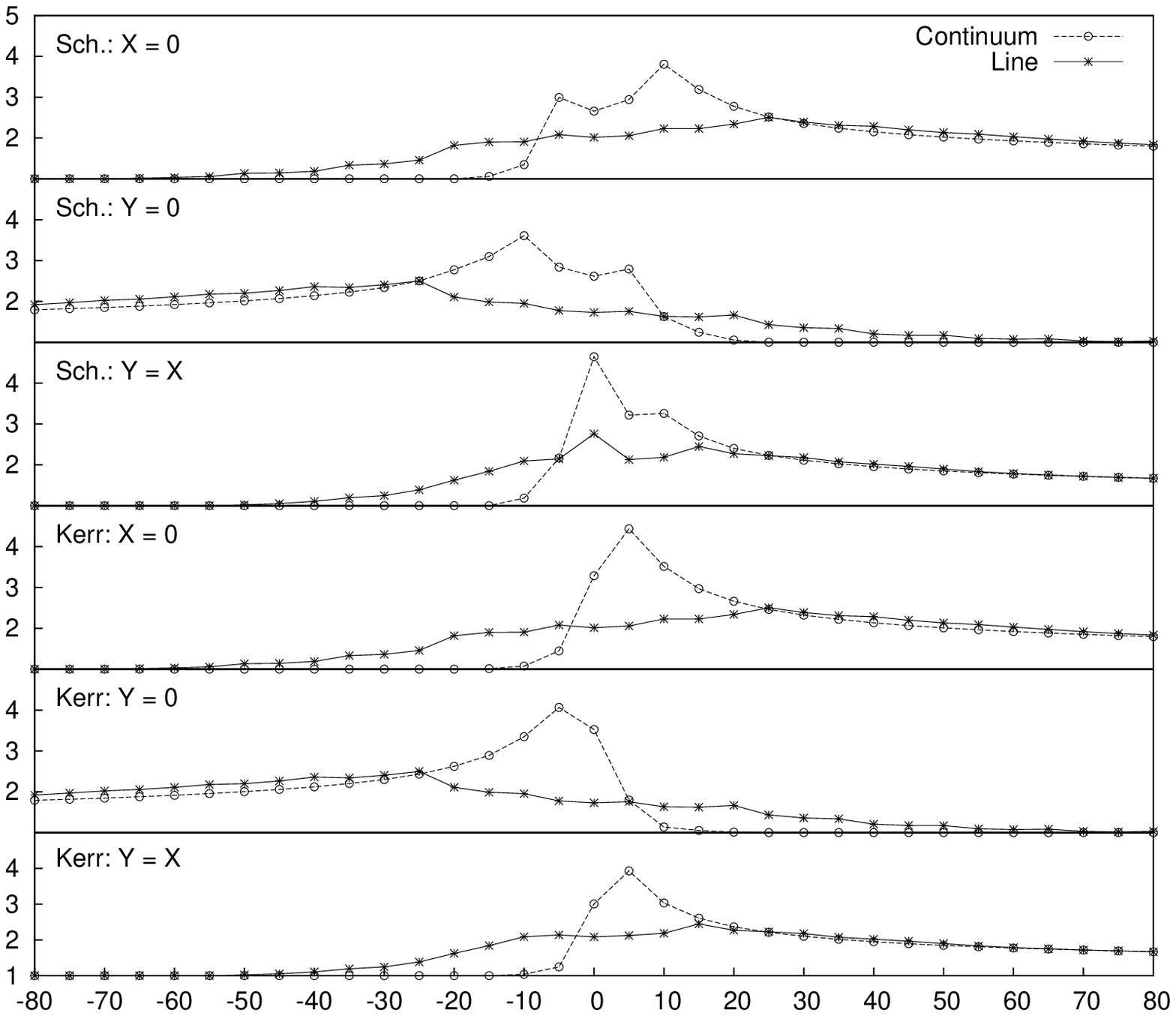}{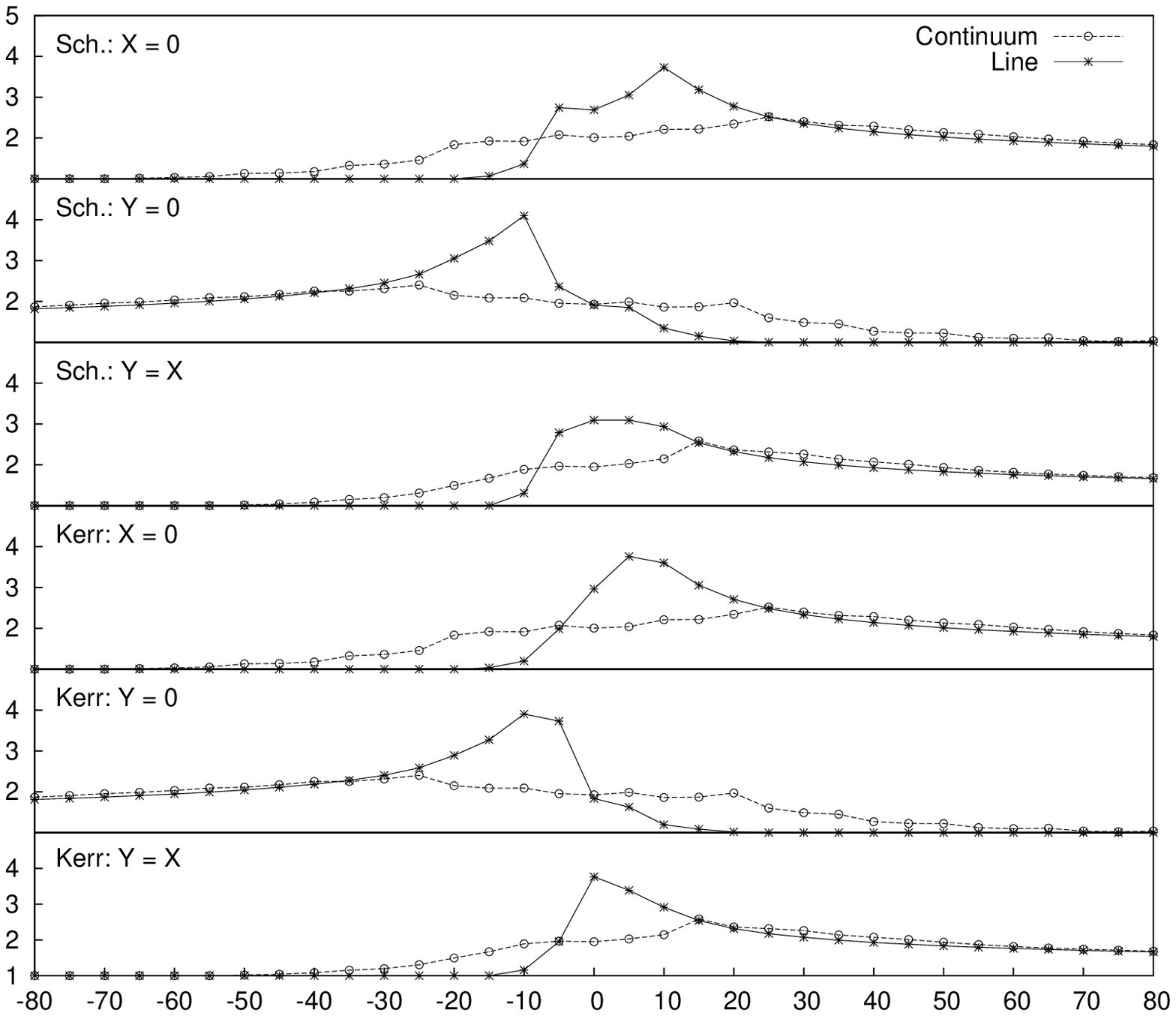}
\caption{The same as in Fig. 5, but for the power-law radiation, the left
panel corresponds
to the case iii) and the right to the case iv) of \S 2.2.1
}
\end{center}
\end{figure}

\begin{figure}
\begin{center}
\plottwo{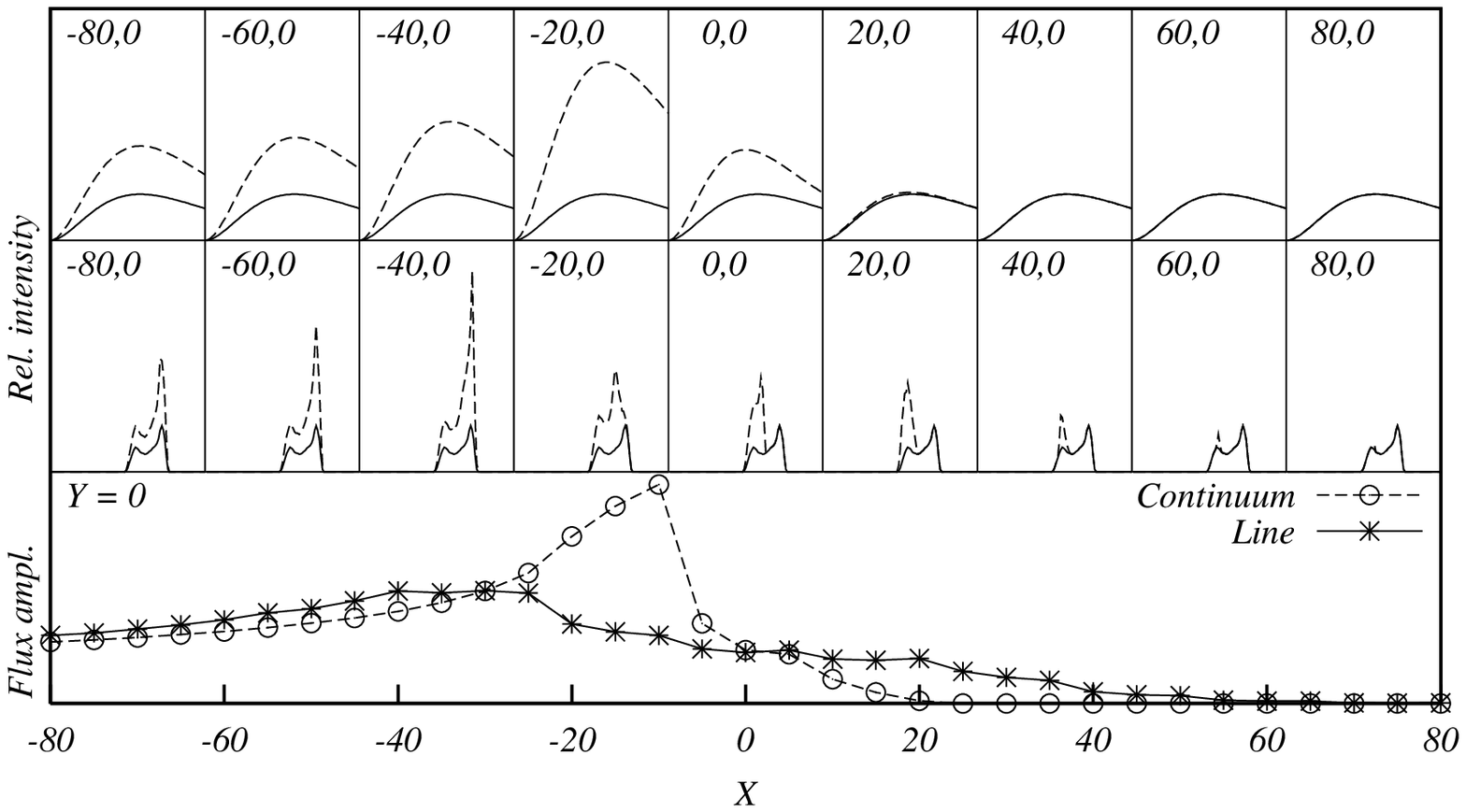}{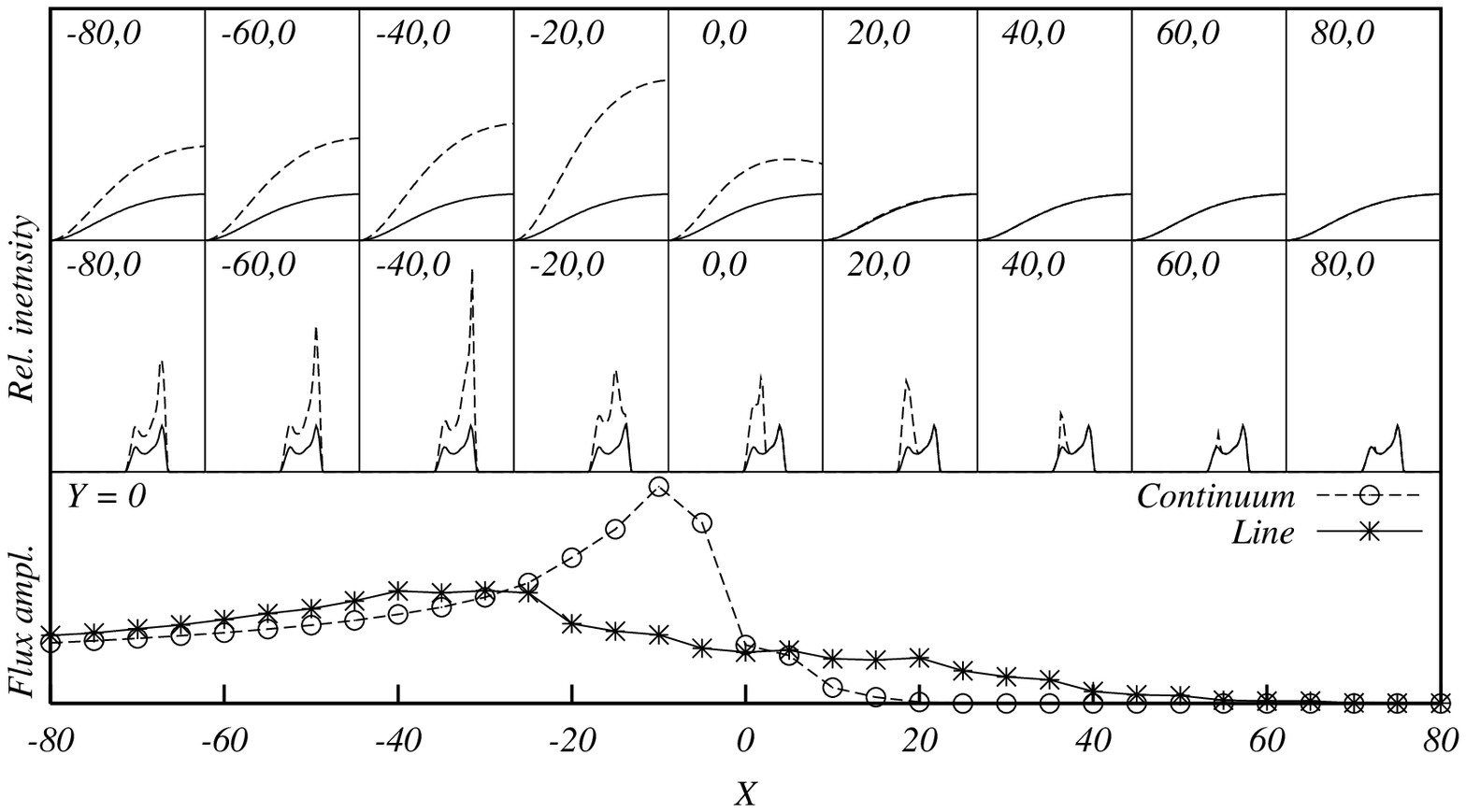}
\caption{Variations of the X-ray continuum and the Fe K$\alpha$ line
flux for a highly inclined disk ($i=75\degr$) due to microlensing by a
caustic with ERR=2000 R$_g$.
 Black body (left) and modified black body
(right) are considered. The geometry corresponds to case (iii) of \S 2.2.1.
The flux axis ranges from 1 to 4.}
\end{center}
\end{figure}
\clearpage

\begin{figure}
\begin{center}
\includegraphics[width=16cm]{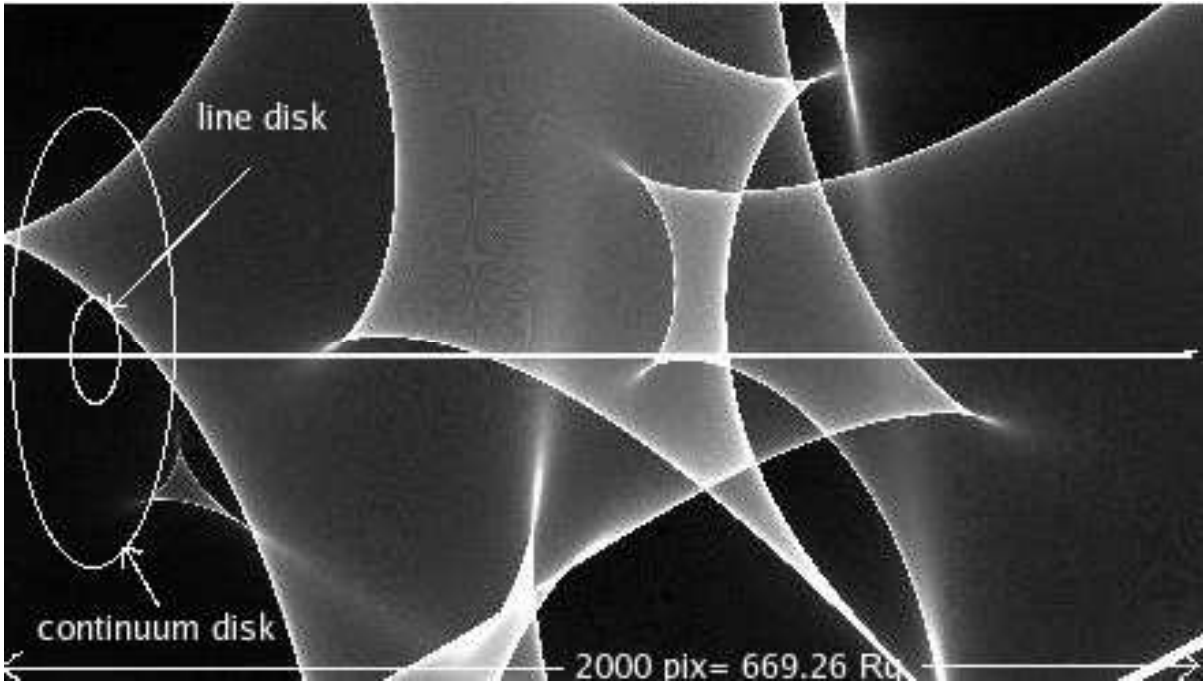}
\caption{Microlensing map of QSO 2237+0305A image with 1ERR$\times$2ERR
(1000 pix $\times$2000 pix=334.63 R$_g\times$669.26R$_g$)
on a side and scheme of the projected disk with outer radius R$_{out}$=20
R$_g$ and
100 R$_g$ for the Fe K$\alpha$ line and the X-ray continuum,
respectively. The straight line presents the path of the center of the
disk (left side of the pattern corresponds to 0 pix).}
\end{center}
\end{figure}

\begin{figure}
\begin{center}
\includegraphics[]{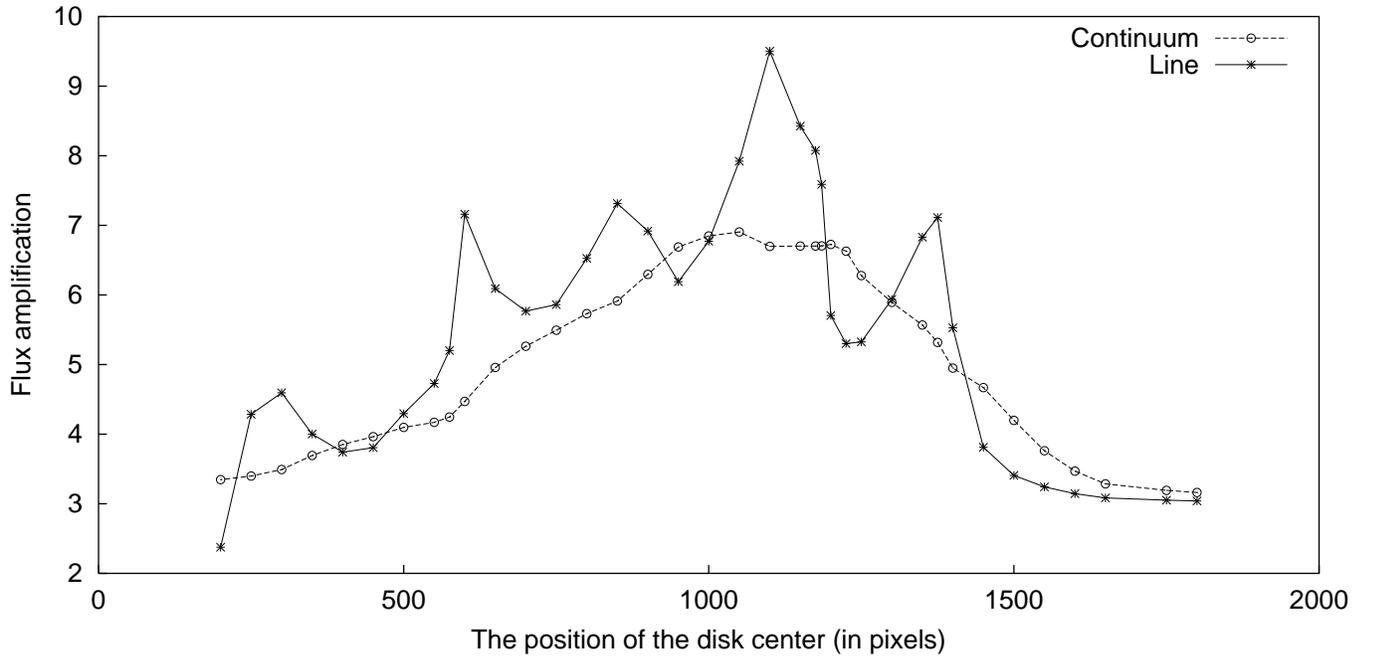}
\caption{The amplification of the Fe K$\alpha$ line and the X-ray
continuum total flux for different positions of the center on the
microlensing map of QSO 2237+0305A image (see Fig. 9).} \end{center}
\end{figure}

\begin{figure}
\begin{center}
\plottwo{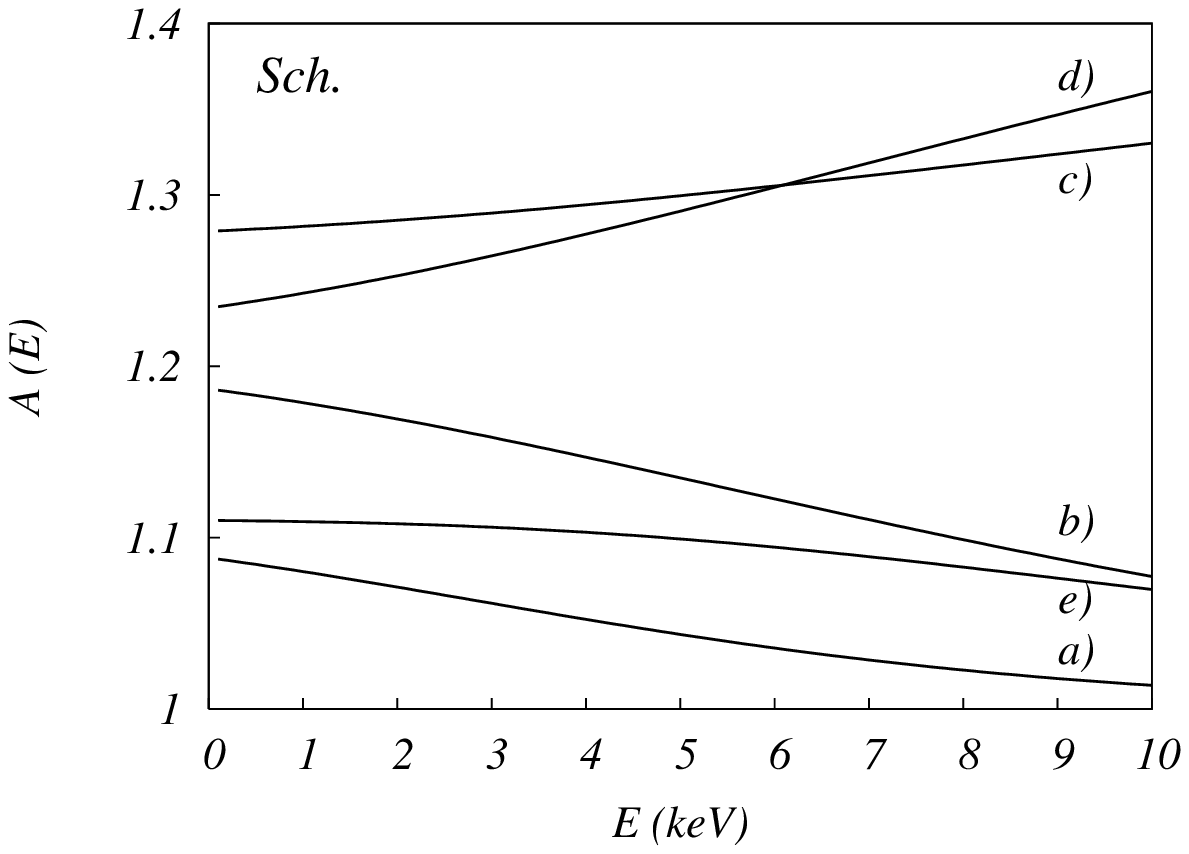}{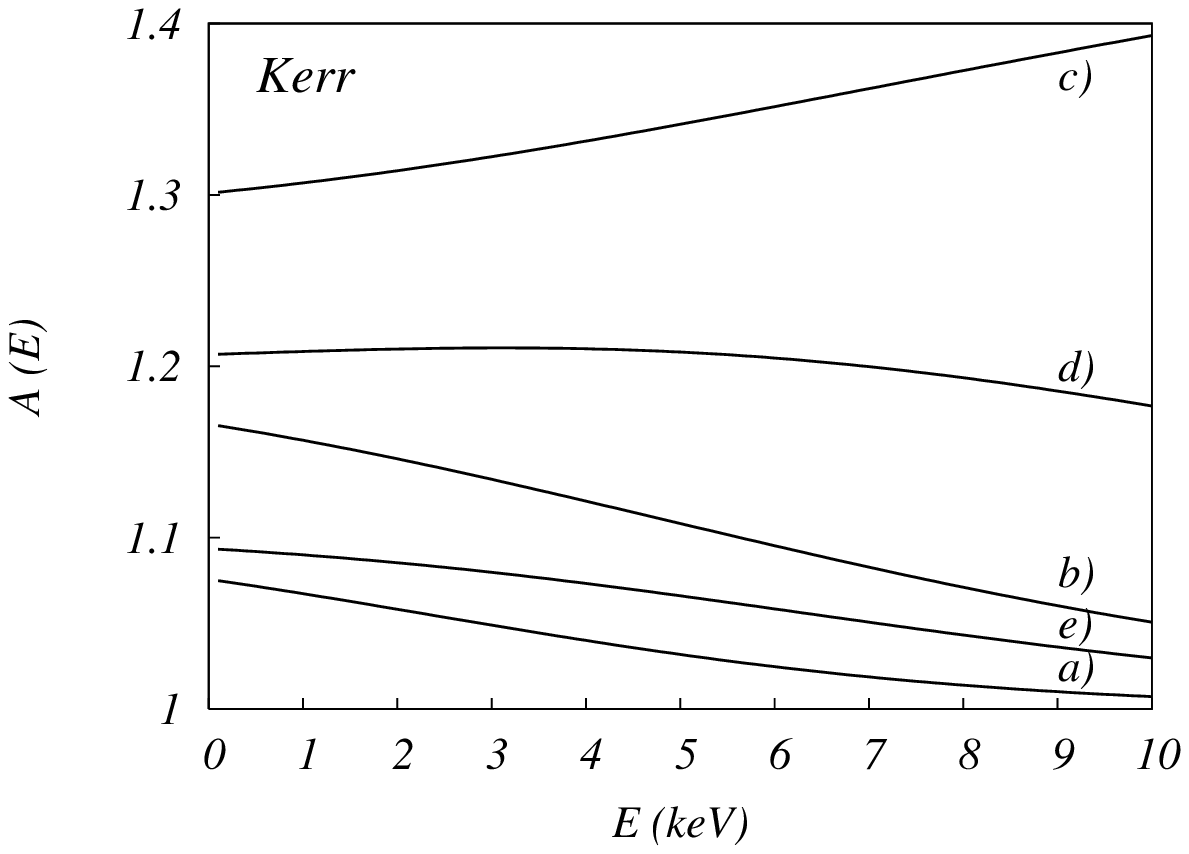}
\caption{Microlensing amplification as a function of emitted energies
(the chromatic effect of microlensing). The calculations were
performed for caustic crossing along the rotation axis (Y=0) for the
following
positions on X-axis: a) X=20 R$_g$; b) X=10 R$_g$; c) X=0 R$_g$; d) X=-10
R$_g$ and e) X=-20 R$_g$.  We take ERR=50 R$_g$. The radii of the
disk are $R_{inn}=R_{ms}$, R$_{out}=30 \ R_g$.  The
 black body emissivity law in both Schwarzschild (left)
and Kerr metrics (right) is used in calculation.}
\end{center}
\end{figure}

\begin{figure}
\begin{center}
\plottwo{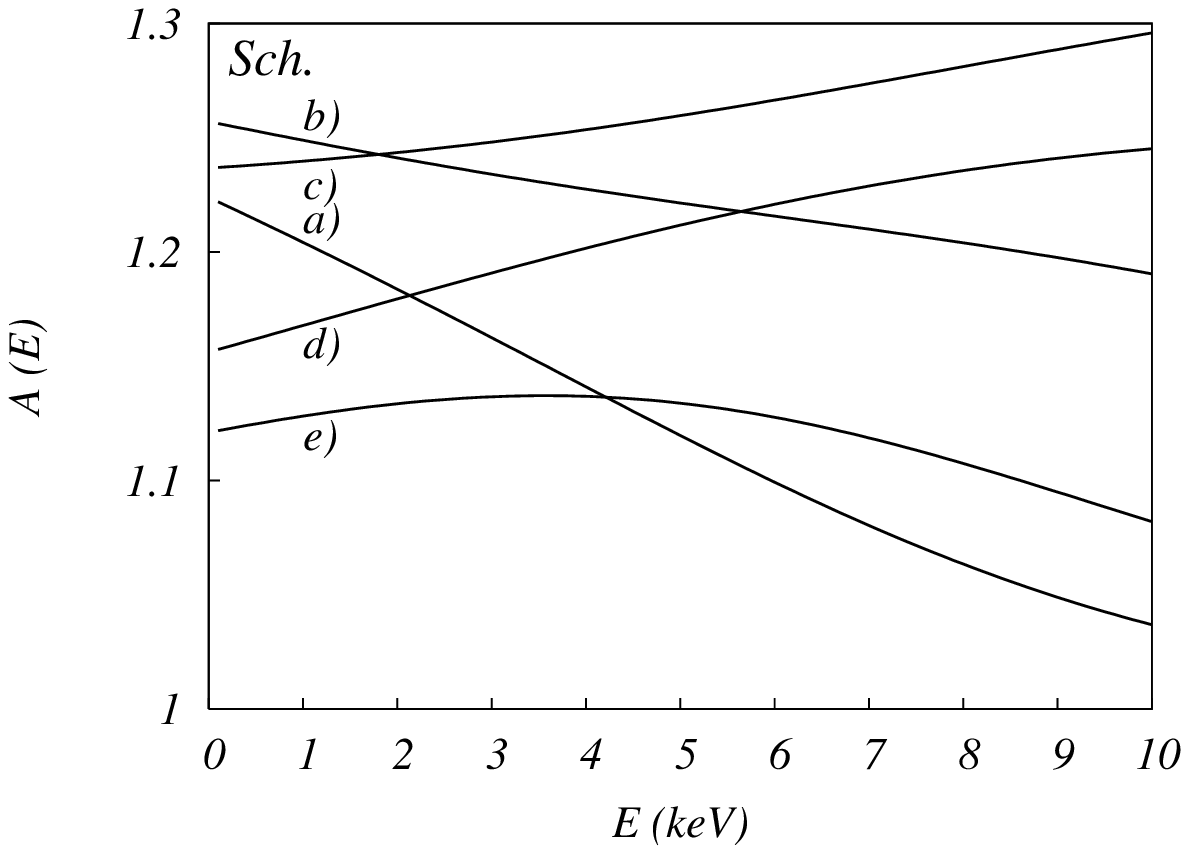}{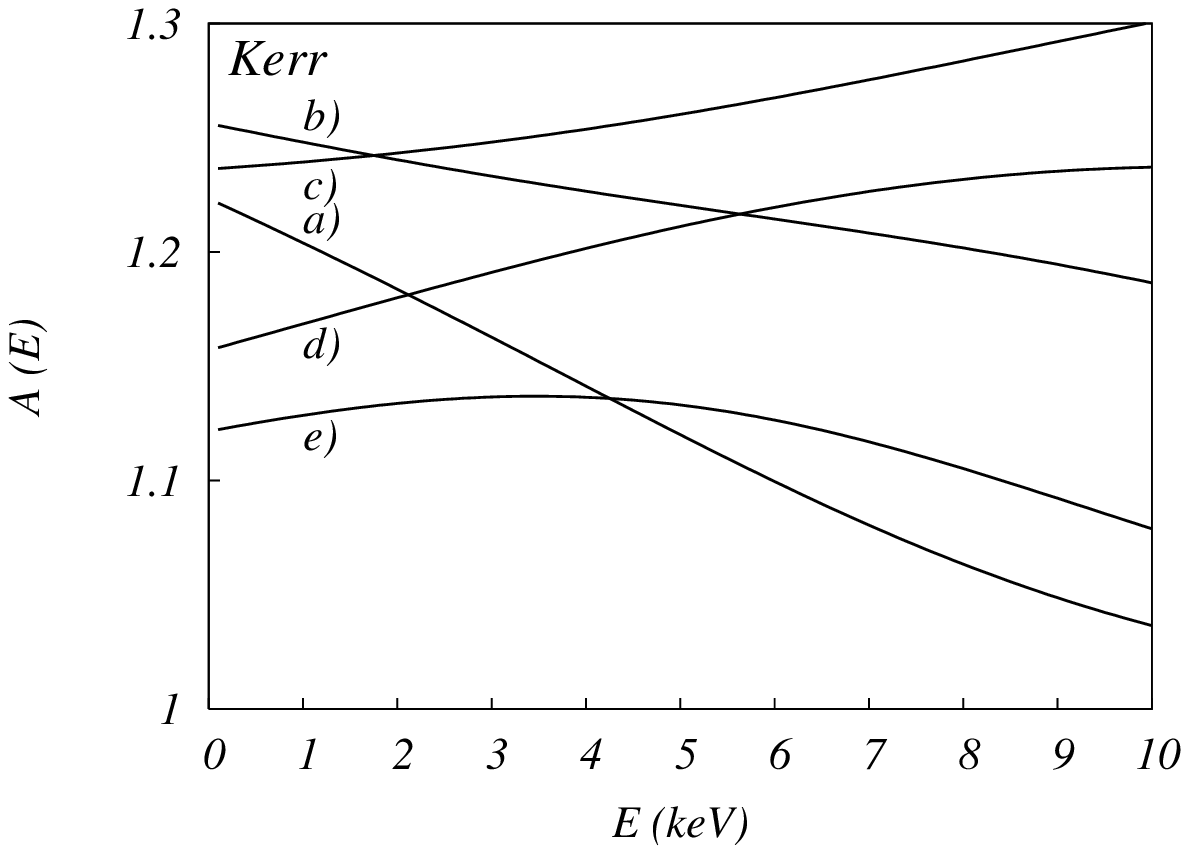}
\caption{The same as in Fig. 11, but for the 'modified' black body law
in Schwarzschild (left) and Kerr metric (right).}
\end{center}
\end{figure}

\clearpage

\begin{table}
\centering
\caption{Projected ERR (expressed in gravitational radii)
  for different deflector masses for the four lensed QSOs where
microlensing of the Fe K$\alpha$ line is suspected: J0414+0534 -
\cite{Chart02a},
  QSO H1413+117 - \cite{Osh01,Chart04}, QSO
2237+0305 - \cite{Dai03}. Used
 values for the cosmological constants are: $H_0=50\rm \ km\
s^{-1}Mpc^{-1}$ and
$\Omega_0=1$. The black hole mass is assumed to be 10$^8\rm M_\odot$.}
\begin{tabular}{|c|c|c|c|c|c|c|c|}
\tableline
 Object & $z_s$ & $z_l$ & $1\times 10^{-4}M_\odot$ &$1\times
 10^{-3}M_\odot$ &$1\times 10^{-2}M_\odot$ &$1\times 10^{-1}M_\odot$ & $1
M_\odot$ \\
\tableline
\tableline
 MG J0414+0534& 2.64  & 0.96  & 20.3   &   64.2   &  203.1  &  642.3  &
2031.1 \\
 QSO 2237+0305& 1.69  & 0.04  & 11.2   &   35.4   &  112.1  &  354.5  &
1121.0 \\
 QSO H1413+117& 2.56  & 1.00  & 19.8   &   62.5   &  197.7  &  625.2  &
1977.0 \\
\tableline
\end{tabular}
\end{table}


\begin{thebibliography}{}

\bibitem[\protect\citeauthoryear{Abajas et al.}{2002}]{Aba02}
Abajas, C., Mediavilla, E.G., Mu\~noz, J.A., Popovi\'c, L. \v C.,
\& Oscoz A. 2002,  ApJ 576, 640.

\bibitem[\protect\citeauthoryear{Abajas et al.}{2004}]{Aba04}
Abajas, C., Mediavilla, E.G., Gil-Merino, R., Mu\~noz, J.A., Popovi\'c,
L. \v C., \& Oscoz A. 2004,  presented in IAU 225 Symposium ``Impact
of Gravitational Lensing on Cosmology'', 19-23 July, Lausanne, Switzerland.

\bibitem[\protect\citeauthoryear{Ballantyne \& Fabian}{2005}]{bf05}
Ballantyne D.R., Fabian, A.C. 2005, ApJ, 622, L97.

\bibitem[\protect\citeauthoryear{Bian \& Zhao}{2002}]{BZ02}
 Bian, W.,
Zhao, Y.  2002, A{\&}A,
395, 465.


\bibitem[\protect\citeauthoryear{Chartas et al.}{2002}]{Chart02a}
Chartas, G., Agol, E., Eracleous, M., Garmire, G., Bautz, M. W., Morgan,
N. D. 2002,  ApJ,  568, 509.

\bibitem[\protect\citeauthoryear{Chartas et al.}{2004}]{Chart04}
Chartas, G., Eracleous, M., Agol, E., Gallagher, S. C. 2004, ApJ,
606, 78.

\bibitem[\protect\citeauthoryear{Dai et al. }{2003}]{Dai03}
Dai, X., Chartas, G., Agol, E., Bautz, M. W., \&
  Garmire, G.P. 2003, ApJ, 589, 100.

\bibitem[\protect\citeauthoryear{Dai et al. }{2004}]{Dai04}
Dai, X., Chartas, G., Eracleous, M. \&
  Garmire, G.P. 2004, ApJ, 605, 45.


\bibitem[\protect\citeauthoryear{Dov\v ciak  et al. }{2004}]{Dov04}
Dov\v ciak, M., Karas, V., Yaqoob, T. 2004, ApJS, 153, 205.

\bibitem[\protect\citeauthoryear{Fabian}{2001}]{Fabian01}
  Fabian, A. 2001,
in Proc. of the 20th Texas Symposium on Relativistic Astrophysics, ed.
  J.~C. Wheeler, H. Martel, AIP Conferences, { 586}, (Americal Institute
of Physics,
  Melville, New York) 643.

\bibitem[\protect\citeauthoryear{Fabian \& Vaughan}{2003}]{Fab03}
Fabian, A.C., Vaughan S. 2003,  MNRAS, 340, L28.

\bibitem[\protect\citeauthoryear{Jaroszy\'nski}{2002}]{Jar02}
Jaroszy\'nski, M. 2002, Acta Astronomica, 52, 203.

\bibitem[\protect\citeauthoryear{Jaroszy\'nski et al.}{1992}]{Jar92}
Jaroszy\'nski, M., Wambsganss, J.W., Paczy\'nski, B. 1992, ApJ, 396, L65.

\bibitem[\protect\citeauthoryear{Kayser et al.}{1986}]{Kay86}
Kayser, R., Refsdal, S., \& Stabell, R. 1986, A{\&}A, 166, 36.


\bibitem[\protect\citeauthoryear{Kochanek}{2004}]{Koch04}
Kochanek, C.S. 2004, ApJ, 605, 58.

\bibitem[\protect\citeauthoryear{Lewis \& Ibata}{2004}]{li04}
Lewis, G. F., Ibata, R.A. 2004, MNRAS, 348,
24.

\bibitem[\protect\citeauthoryear{Mineshige et al.}{2001}]{Min01}
Mineshige, S., Yonehara, A., \& Takahashi, R. 2001,
PASA, 18, 186.

\bibitem[\protect\citeauthoryear{Nandra et al.}{1997}]{Nan97}
 Nandra K., George I.M., Mushotzky R.F., Turner T.J. \&
Yaqoob T. 1997, ApJ. 477, 602.

\bibitem[\protect\citeauthoryear{Nandra et al.}{1999}]{Nan99}
Nandra, K., George, I. M., Mushotzky, R. F., Turner, T. J., Yaqoob,
T. 1999, ApJ 523, 17.

\bibitem[\protect\citeauthoryear{Novikov \& Thorne}{1973}]{Novikov73}
Novikov, I.D., \& Thorne, K.S.  1973, Black Holes (Eds. C. De Witt,
B. De Witt), Gordon \& Breach, 344.

\bibitem[\protect\citeauthoryear{Oshima et al.}{2001a}]{Osh01}
Oshima, T., Mitsuda, K., Fujimoto R., Iyomoto N., Futamoto K., et
al. 2001a,   ApJ,  563, L103.

\bibitem[\protect\citeauthoryear{Oshima et al.}{2001b}]{Osh02}
Oshima, T., Mitsuda, K., Ota, N., Yonehara, A., Hattori, M., Mihara, T.
\& Sekimoto, Y. 2001b, ApJ, 551, 929.

\bibitem[\protect\citeauthoryear{Page et al.}{2004}]{Pag04}
Page, K.L., Reeves, J.N., O'Brien, P.T., Turner, M.J.L., Worrall,
D.M. 2004, MNRAS, 353, 133.

\bibitem[\protect\citeauthoryear{Popovi{\'c} \& Chartas}{2005}]{Pc04}
Popovi\'c, L. \v C., Chartas, G. 2005, MNRAS, 357, 135.

\bibitem[\protect\citeauthoryear{Popovi{\'c} et al.}{2001a}]{Popov01}
Popovi{\'c}, L., \v C.,  Mediavilla, E.G., Mu\~noz J.,
Dimitrijevi\'c, M.S., \& Jovanovi\'c, P. 2001a,  SerAJ,
164, 73.

\bibitem[\protect\citeauthoryear{Popovi{\'c} et al.}{2001b}]{Pop01b}
Popovi{\'c}, L.\v C.,  Mediavilla, E.G., \& Mu\~noz J. 2001b,
A{\&}A 378, 295.

\bibitem[\protect\citeauthoryear{Popovi{\'c} et al.}{2003a}]{Pop03}
 Popovi\'c, L.\v C., Mediavilla, E.G., Jovanovi\'c, P., \& Mu\~noz, J.A.
 2003a,  A{\&}A, 398, 975.

\bibitem[\protect\citeauthoryear{Popovi{\'c} et al.}{2003b}]{Pop03a}
Popovi\'c, L.\v C., Jovanovi\'c, P., Mediavilla, E.G., \& Mu\~noz,
J.A. 2003b, Astron.  Astrophys. Transactions, 22, 719.

\bibitem[\protect\citeauthoryear{Richards et al.}{2004}]{Ric04}
Richards, G.T., Keeton, C.R., Pindor, B. et al. 2004, ApJ, 610, 679.

\bibitem[\protect\citeauthoryear{Schneider \& Weiss}{1987}]{Sch87}
Schneider, P., \& Weiss, A. 1987, A{\&}A, 171, 49.


\bibitem[\protect\citeauthoryear{Shakura \& Sunyaev}{1973}]{Shakura73}
 Shakura, N.I., \& Sunyaev, R.A.
 1973,  A{\&}A, 24, 337.


\bibitem[\protect\citeauthoryear{Shalyapin et al.}{2002}]{Sha02}
Shalyapin, V., N., Goicoechea, L., J., Alcalde, D., Mediavilla
 E., Mu\~noz J.A., Gil-Merino, R.  2002, ApJ, 579,
127.

\bibitem[\protect\citeauthoryear{Shapiro \& Teukolsky}{1983}]{Shapiro83}
 Shapiro, S.L., \& Teukolsky, S.A. 1983,
Black Holes,  White Dwarfs and Neutron Stars: Physics of Compact
Objects, John Wiley \& Sons,  New York.


\bibitem[\protect\citeauthoryear{Straumann}{1984}]{Straumann84}
 Straumann, N. 1984,
General Relativity and Relativistic Astrophysics, Springer-Verlag,
Berlin, Heidelberg, New York.

\bibitem[\protect\citeauthoryear{Takahashi et al.}{2001}]{Tak01}
Takahashi, R., Yonehara, A., Mineshige, S. 2001, Publ. Astron. Soc. Japan
53, 387.

\bibitem[\protect\citeauthoryear{Wang et al.}{2003}]{Wang03}
Wang, J.-M., Ho, L. C.; Staubert, R. 2003, A{\&}A, 409, 887.

\bibitem[\protect\citeauthoryear{Wambsganss \& Paczynski}{1991}]{WP91}
Wambsganss, J., Paczynski, B. 1991, AJ, 102, 86.

\bibitem[\protect\citeauthoryear{Wambsganss et al.}{1990a}]{WP90}
Wambsganss, J., Paczynski, B., \& Katz, N. 1990a, ApJ, 352, 407.

\bibitem[\protect\citeauthoryear{Wambsganss et al.}{1990b}]{WSP90}
Wambsganss, J., Schneider, P., \& Paczynski, B. 1990b, ApJ, 358, L33.

\bibitem[\protect\citeauthoryear{Witt et al.}{1993}]{Witt03} Witt H.J., Kayser R., Refsdal S., 1993, A{\&}A 268, 501.

\bibitem[\protect\citeauthoryear{Wyithe et al.}{2000}]{Wy00}
Wyithe, J. S. B., Webster, R. L., Turner, E. L., Agol, E. 2000, MNRAS,
318, 1105.

\bibitem[\protect\citeauthoryear{Yonehara et al.}{1999}]{Yon99}
Yonehara, A., Mineshige, S.,  Fukue, J., Umemura, M., Turner,
E.L. 1999,
A{\&}A, 343, 41.

\bibitem[\protect\citeauthoryear{Yonehara et al.}{1998}]{Yon98}
Yonehara, A., Mineshige, S., Manmoto, T., Fukue, J., Umemura, M., Turner,
E.L. 1998,
ApJ, 501, L41; ApJ, 511, L65.


\bibitem[\protect\citeauthoryear{Zakharov et al.}{2003}]{ZKLR02}
  Zakharov, A.F.,  Kardashev, N.S., Lukash, V.N., \& Repin, S.V. 2003,
MNRAS, 342, 1325.

\bibitem[\protect\citeauthoryear{Zakharov \& Repin}{2002a}]{Zak_rep02}
  Zakharov, A.F., \& Repin, S.V. 2002a, Astronomy Reports, 46, 360.

\bibitem[\protect\citeauthoryear{Zakharov \& Repin}{2002b}]{Zak_rep02a}
  Zakharov, A.F., \& Repin, S.V. 2002b,
in Proc. of the Eleven Workshop
  on General Relativity  and Gravitation in Japan, ed.
  J.~Koga, T.~Nakamura, K.~Maeda, K.~Tomita, (Waseda University,
  Tokyo) 68.

\bibitem[\protect\citeauthoryear{Zakharov \&
Repin}{2002c}]{Zak_rep02_xeus}
  Zakharov, A.F., \& Repin, S.V. 2002c, in
  Proc. of the Workshop "XEUS - studying the evolution of the hot
  Universe", ed. G. Hasinger, Th. Boller, A.N. Parmar,  MPE Report~281,
339.

\bibitem[\protect\citeauthoryear{Zakharov \&
Repin}{2003a}]{Zak_rep03a_AA}
  Zakharov, A.F., \& Repin, S.V. 2003a, Nuovo Cim., 118B, 1193.


\bibitem[\protect\citeauthoryear{Zakharov \&
Repin}{2003b}]{Zak_rep03b_AA}
  Zakharov, A.F., \& Repin, S.V. 2003b, A{\&}A, 406,7.


\bibitem[\protect\citeauthoryear{Zakharov et al.}{2004}]{Zak03}
Zakharov, A.F., Popovi\'c, L. \v C., Jovanovi\'c, P. : 2004,
 A{\&}A, 420, 881.

\end{thebibliography}
\end{document}